\documentclass[a4paper,11pt]{article}

\usepackage[normalem]{ulem}

\usepackage{xcolor}
\usepackage[english]{babel}
\usepackage[T1]{fontenc}
\usepackage[utf8]{inputenc}
\usepackage{authblk}
\usepackage{mathtools}
\usepackage{epsfig}
\usepackage{slashed}
\usepackage{amsmath,amssymb}
\usepackage{mathrsfs}
\usepackage{amsfonts}
\usepackage{enumitem}
\usepackage{graphicx,color,xcolor}
\usepackage{cite}
\usepackage{float}
\usepackage{subcaption}

\usepackage{hyperref}

\hypersetup{
	colorlinks=true,
	linkcolor=blue,
	filecolor=magenta,      
	urlcolor=cyan,
}
\usepackage{cleveref}
\usepackage[left=2.5cm,right=2.5cm,top=2.5cm,bottom=2.5cm]{geometry}
\numberwithin{equation}{section} 

\definecolor{refcol}{rgb}{0.9,0.1,0.1}
\hypersetup{colorlinks=true,linkcolor=blue,citecolor=refcol,urlcolor=cyan,linktocpage}

\graphicspath{{Plots/}}

\begin{document}
\begin{titlepage}
\thispagestyle{empty}
\title{
{\Huge\bf Krylov complexity of deformed conformal field theories}}
\vfill
\author{{\bf Arghya Chattopadhyay$^a$}\thanks{{\tt \href{mailto:arghya.chattopadhyay@umons.ac.be}{arghya.chattopadhyay@umons.ac.be}}}, {\bf Vinay Malvimat$^b$}\thanks{{\tt \href{mailto:vinay.malvimat@apctp.org}{vinay.malvimat@apctp.org}}},
{\bf Arpita Mitra$^c$}\thanks{{\tt \href{mailto:arpitamitra89@gmail.com}{arpitam@postech.ac.kr}}}
\smallskip\hfill\\      	
\small{$^a${\it Service de Physique de l'Univers, Champs et Gravitation, Université de Mons}\\{\it 20 Place du Parc, 7000 Mons, Belgium.}\\\hfill\\
 $^b${\it Asia Pacific Center for Theoretical Physics, 77 Cheongam-ro, Nam-gu, Pohang-si}\\
{\it Gyeongsangbuk-do, 37673, Korea.}\\
\hfill\\
$^c${\it Department  of  Physics,  Pohang  University  of  Science  and  Technology}\\{\it  Pohang  37673,  Korea.}\\}}
\vfill
\date{\begin{quote}
\centerline{{\bf Abstract}}\vspace{1em}
{\small We consider a perturbative expansion of the Lanczos coefficients and the Krylov complexity for two-dimensional conformal field theories under integrable deformations. Specifically, we explore  the consequences of T$\bar{\text{T}}$, J$\bar{\text{T}}$, and J$\bar{\text{J}}$ deformations, focusing on first-order corrections in the deformation parameter. Under T$\bar{\text{T}}$ deformation, we demonstrate that the Lanczos coefficients $b_n$ exhibit unexpected behavior, deviating from linear growth within the valid perturbative regime. Notably, the Krylov exponent characterizing the rate of exponential growth of complexity surpasses that of the undeformed theory for positive value of deformation parameter, suggesting a potential violation of the conjectured operator growth bound within the realm of perturbative analysis. One may attribute this to the existence of logarithmic branch points along with higher order poles in the autocorrelation function compared to the undeformed case. In contrast to this, both J$\bar{\text{J}}$ and J$\bar{\text{T}}$ deformations induce no first order correction to either the linear growth of  Lanczos coefficients at large-$n$  or the Krylov exponent  and hence the results for these two deformations align with those of the undeformed theory.}
\end{quote}}
\end{titlepage}
\thispagestyle{empty}\maketitle\vfill \eject

\tableofcontents

\section{Introduction}\label{sec:intro}
The concept of quantum complexity originally emerged as a measure of algorithmic efficiency in quantum computation. It involved assigning distinct costs to various configurations of quantum gates for transitioning from a specified reference state to a desired target state. While Nielsen complexity of a unitary operator is defined as the minimal distance of it from the identity in the group manifold \cite{Nielsen_1,Nielsen_2006,Nielsen_2,Nielsen_3}, the computational complexity counts the minimum number of elementary quantum gates necessary
to achieve a unitary transformation to reach the target state from the reference state. In recent years, there has been significant interests in assigning cost for preparing a state within holography, spurred by Susskind's conjecture linking {the notion of} cost assignment to black hole physics \cite{Susskind:2014rva}. Due to several universal features exhibited by computational complexity in chaotic systems of finite size, \cite{Susskind:2014rva} conjectured its relations with geometric aspects of anti-de Sitter (AdS) black hole geometries, particularly concerning the expansion of black hole interiors over time. Among the most extensively examined proposals concerning holographic complexity are the ``complexity=volume'' \cite{Susskind:2014rva, Stanford:2014jda}, ``complexity=action'' \cite{Brown:2015bva, Brown:2015lvg} and ``complexity=volume 2.0'' \cite{Couch:2016exn} proposals\footnote{Several works have explored these propositions \cite{Carmi:2017jqz, Alishahiha:2015rta, Alishahiha:2017hwg, Alishahiha:2018swh, Chapman:2017rqy, Chapman:2018hou, Chapman:2021jbh, Auzzi:2019fnp,Auzzi:2021ozb, Omidi:2022whq, Bhattacharya:2023drv} and also examined the violation of universal bounds within this framework \cite{Alishahiha:2018tep, Mandal:2022ztj}, for instance, the Lloyd's bound \cite{Aguilar-Gutierrez:2023ccv}.}. Recently, a broad array of gravitational observables has been introduced as ``complexity=anything'' \cite{Belin:2021bga, Belin:2022xmt, Myers:2024vve}. The endeavor to comprehend holographic complexity through a dual microscopic description has motivated advancements in defining and examining complexity within quantum field theories.  However the relationship to holographic complexity has largely been constrained to qualitative comparisons. For example, the time evolution of holographic complexity exhibits a linear growth up to saturating to almost a constant value, similar to the computational complexity. In \cite{Erdmenger:2022lov}, a mapping between computational complexity and a geometric entity within the gravity theory was derived from first principles and can be naturally applied to AdS black hole spacetimes and their dual thermofield double (TFD) states.\vspace{1em}\\
Gradually complexity has evolved into a crucial tool to probe quantum chaos \cite{Ali:2019zcj,Bhattacharyya:2019txx,Bhattacharyya:2020iic,Bhattacharyya:2020art,Balasubramanian:2021mxo,Balasubramanian:2022tpr}, motivated by the expectation that chaotic systems exhibit maximum complexity. In parallel to this developments, there exist several other approaches to diagnose quantum chaos including level spacing statistics \cite{Pal:2010bew, Russomanno:2020fnj}, exponential growth of out-of-time-order correlators (OTOCs) \cite{Maldacena:2015waa}, eigenstate thermalization hypothesis \cite{Deutsch:1991msp, Srednicki:1994mfb, Deutsch:2018ulr}. One of the significant challenges in this regard is to precisely examine if there exists a notion which unifies these varied approaches. Recently, a new measure of quantum state/operator complexity known as the Krylov complexity has been introduced in \cite{Parker:2018yvk} to analyze chaotic behaviour of quantum systems. Krylov complexity essentially characterizes the spread of an operator or equivalently a state evolving with time in a distinct Hilbert space {spanned by what is} known as the Krylov basis\footnote{Krylov state complexity is also known as spread complexity in the literature.}. This measure of complexity has more universal feature compared to existing ones, since for a maximally entangled state the Krylov complexity solely relies on the Hamiltonian's spectrum and does not depend on the choice of fundamental gates or any local information unlike both the Nielsen and computational complexity. Time evolution of an unitary operator in a space of operators in the Heisenberg picture can be described as the evolution under the operation of a super operator known as the Liouvillian whose repeated action leads to nested commutators involving the Hamiltonian and the operator itself. This in turn leads to the definition of an invariant subspace known as the Krylov subspace spanned by orthonormalized operators constructed from iterative action of the Liouvillian \cite{Viswanath:2008, Parker:2018yvk}. \vspace{1em}

\noindent It was argued in \cite{Parker:2018yvk}, that for maximally chaotic systems the Lanczos coefficients exhibit a linear growth (maximally allowed growth) with the number of basis vectors (up to a log correction) satisfying an universal operator growth hypothesis. While the OTOC typically exhibits an exponential growth over time, indicating a rapid scrambling of the information  and displaying an extreme sensitivity to the initial conditions, the Krylov complexity also grows exponentially fast as $e^{\lambda_K t}${, with $\lambda_K$ being the Krylov exponent and $t$ is time}. Furthermore it was conjectured in \cite{Parker:2018yvk}, that twice the rate of linear growth of Lanczos ( which is same as the Krylov exponent )  serves as an upper bound on the quantum Lyapunov exponent $\lambda_L$ derived from OTOCs $\lambda_{L}\leq \lambda_K$\footnote{The relationship that twice the linear growth rate of the Lanczos coefficients equals the Krylov exponent can be demonstrated using a continuum version of the recursion equation for operator amplitudes \cite{Barbon:2019wsy}. }.  Since quantum Lyapunov exponent $\lambda_L$ itself satisfies the {Maldacena-Shenker-Stanford (MSS)} bound  $\lambda_L\leq \frac{2\pi}{\beta}$ \cite{Maldacena:2015waa}, combining the two bounds a modified bound $\lambda_{L}\leq \lambda_{K}\leq \frac{2\pi}{\beta}$ was conjectured to hold for any quantum system in \cite{Avdoshkin:2022xuw} (from now on we will refer to this as the extended MSS bound). Although the operator growth hypothesis has turned out to be very efficient in distinguishing chaotic and non-chaotic systems in lattice models, it was {later shown to be}  far too universal in quantum systems with infinite degrees of freedom such as QFTs. Infact even in very simple  integrable field theories such as free scalar field theories the Lanczos coefficients exhibit linear growth and the Krylov complexity also grows exponentially \cite{Dymarsky:2019elm, Dymarsky:2021bjq}.  Furthermore, in the same article it was also shown a primary operator in any two dimensional CFT also exhibits a linear growth of Lanczos coefficients leading to exponential growth of Krylov complexity with Krylov exponent saturating the conjectured bound.  This indicates that the notion of the Krylov complexity is more subtle and its behaviour might in turn be state dependent through the choice of the inner product in the Krylov space as shown in\cite{Kundu:2023hbk}.\vspace{1em}  

\noindent Due to many intriguing features such as the ones mentioned above, the behaviour of  Krylov complexity has been extensively investigated in a vast number of different settings such as  free and interacting {Quantum Field Theories}(QFTs), random matrix theories, open quantum systems etc \cite{Dymarsky:2019elm, Dymarsky:2021bjq, Camargo:2022rnt, Banerjee:2022ime,Iizuka:2023pov, Iizuka:2023fba,Balasubramanian:2022tpr, Erdmenger:2023wjg,Bhattacharjee:2022ave, Bhattacharjee:2023uwx, Bhattacharya:2023zqt, Bhattacharyya:2023grv, Beetar:2023mfn, Camargo:2023eev, Malvimat:2024vhr, Caputa:2024vrn,Afrasiar:2022efk}. Another interesting behavior exhibited by the Krylov complexity of quantum chaotic systems, encompassing notable models such as the SYK model and matrix models reveals a fascinating pattern wherein complexity exhibits four distinct dynamical phases: a steadily rising ramp culminating in a peak, succeeded by a declining slope leading to a plateau \cite{Balasubramanian:2022tpr}. Notably, the duration of these phases, as well as the magnitudes of the peak and plateau, exhibit exponential dependence on entropy. This observation draws parallel to the characteristic slope-dip-ramp-plateau structure observed in the spectral form factor \cite{Guhr:1997ve, Brezin:1997rze, Cotler:2016fpe}, offering valuable insights into the intricate dynamics of chaotic systems. These phases were also observed in integrable systems \cite{Erdmenger:2023wjg, Huh:2023jxt} which exhibits saddle-dominated scrambling. Apart from the above described properties, the notion of Krylov complexity also seems to possess interesting geometrical aspects. Remarkably for low rank algebras, Krylov operator demonstrates a connection to the vector fields responsible for generating isometries within the classical phase space geometry which in turn is also related to Nielsen complexity \cite{Chattopadhyay:2023fob, Craps:2023ivc,Aguilar-Gutierrez:2023nyk}. Its expectation value corresponds to the volume delineated by the classical motion unfolding within the associated geometry \cite{Magan:2018nmu, Chattopadhyay:2023fob}.\vspace{1em}\\
In this work, we delve into the behaviour of Krylov complexity for two dimensional conformal field theories in the presence of perturbative deformations. In particular, we will consider two irrelevant deformations induced by composite operators constructed from abelian current and energy momentum tensor namely T$\bar{\text{T}}$ and J$\bar{\text{T}}$ along with one exactly marginal deformation J$\bar{\text{J}}$ constructed only from abelian currents \cite{Smirnov:2016lqw, Frolov:2019xzi, Cardy:2019qao}. We first compute the thermal autocorrelation functions from the first order correction of the correlation functions due to the deformations. Following this, we determine the first order correction to the Lanczos coefficients and operator wavefunctions due to the deformations for both the positive and negative values of the deformation parameter. Furthermore, finally we examine the growth of Krylov complexity to understand the implications of the above mentioned deformations. The key insight of our study requires to understand the regime of validity of perturbative techniques while computing the Lanczos coefficients and Krylov complexity growth. It is reasonable to expect that perturbation techniques will remain valid as long as the probability of the operator wavefunction remains conserved over time \cite{Bhattacharjee:2022ave}. In \cite{Heveling:2022orr}, stability of autocorrelation function was examined under a specific perturbations to Lanczos coefficients. {Our results demonstrate a correction to the Lanczos coefficients, operator wavefunctions and the growth of Krylov complexity for the T$\bar{\text{T}}$ deformations treated perturbatively, while J$\bar{\text{J}}$ and J$\bar{\text{T}}$ deformation do not affect the CFT results.} This may establish an intriguing behavior of complexity of field theories with the presence of irrelevant deformations, since Lyapunov exponent computed from OTOC do not receive any correction for perturbative T$\bar{\text{T}}$ and J$\bar{\text{T}}$ deformations \cite{He:2019vzf}.\vspace{1em}\\
This paper is organized as the following. We begin by reviewing some properties of T$\bar{\text{T}}$, J$\bar{\text{T}}$ and J$\bar{\text{J}}$ deformations to 2D CFT in \cref{Sec:basics}, focusing mainly on the two point correlation functions within the perturbative framework. In {\cref{Sec:main}}, we provide the main results containing the corrections to the autocorrelation functions, Lanczos coefficients, operator wavefunctions and growth of Krylov complexity due to the perturbative deformations of 2D CFTs, following a brief review on the computation of Krylov complexity. We conclude in \cref{Sec:conclusion}, with possible outlooks and future directions. We provide all the details concerning the integrals required for the derivations of correlation functions in the appendix \ref{app:integrals}.
\section{Brief review of perturbative T$\bar{\text{T}}$, J$\bar{\text{T}}$ and J$\bar{\text{J}}$ deformations to 2D CFT}\label{Sec:basics}
The space of QFTs can be navigated through the renormalization group (RG) flows, wherein conformal field theories are the fixed points. Investigating flows diverging from fixed points proves equally intriguing. These deformations of QFTs, contingent upon the operators driving the flow, fall into three broad categories: relevant, marginal, and irrelevant. Although, irrelevant deformations of QFTs inherently entail an infinite array of counter terms in the Lagrangian, however, recent revelations indicate that for certain specialized classes of irrelevant deformations within two-dimensional spacetime, this process is more tractable and even solvable. In
\cite{Smirnov:2016lqw, Cavaglia:2016oda}, an irrelevant deformation induced by T$\bar{\text{T}}$ composite operator\footnote{In our case with CFT, T$\bar{\text{T}}=4\pi^2 (T_{zz}T_{\bar{z}\bar{z}}-T^2_{z\bar{z}})=T\bar{T}$ where $T_{zz}$ and $T_{\bar{z}\bar{z}}$ are the holomorphic and antiholomorphic component of the CFT stress tensor and $T_{z\bar{z}}=0$.} constructed from the components of the conserved energy momentum tensor was introduced. Given a one parameter family of 2D CFTs, the action of deformed theory along the flow satisfies,
\begin{equation}
 \frac{\partial S}{\partial \lambda}=\int \sqrt{g}~d^2x (\text{T}\bar{\text{T}})_{\lambda}   
\end{equation}
where $\lambda$ is the coupling parameter. Despite the fact that this deformation flows toward the ultraviolet (UV) regime, numerous physically significant quantities like deformed S matrix \cite{Dubovsky:2017cnj}, finite volume spectrum using  integrability methods, torus partition function \cite{Cardy:2018sdv, Datta:2018thy, Aharony:2018bad, Tian:2024vln} could be computed exactly. This deformation is integrable and one can {show that for integrable field theories,} infinite number of conserved charges continue to remain conserved along the flow \cite{Smirnov:2016lqw}. Since most QFTs are not integrable, but the surprising solvability of T$\bar{\text{T}}$ even in that case leads to the random geometry picture prescribed in \cite{Cardy:2018sdv}. Another interesting property of this deformation to a QFT is that it is equivalent to coupling the theory to two dimensional Jackiw-Teitelboim (JT) gravity \cite{Dubovsky:2017cnj}. The holographic dual of this deformed CFT is a cut-off AdS gravity theory \cite{McGough:2016lol,Kraus:2018xrn, Jafari:2019qns}.\vspace{1em}

\noindent For our purpose of computation of Krylov complexity, we are interested in the two point correlation function defined within the perturbative framework. Now we may express the deformed action at leading order as,
\begin{align}
    S(\lambda)=S_0+\lambda \int \sqrt{g}~d^2x (\text{T}\bar{\text{T}})_{0} +\mathcal{O}(\lambda^2)
\end{align}
{where $S_0$ is the undeformed action and $(\text{T}\bar{\text{T}})_{0}$ is the perturbing operator of a CFT.}
For a two dimensional  Euclidean CFT on a complex plane whose coordinates are denoted by ($z, \bar{z}$), the first order correction to a two point function involving two primary operators can be computed following the OPE of energy momentum tensor and subsequently utilizing the Ward identity associated with the conserved energy momentum tensor  as follows\cite{DiFrancesco:1997nk, He:2019vzf, He:2023kgq},
\begin{align}
  \langle O(z_1,\bar{z}_1)O(z_2,\bar{z}_2)\rangle_{\lambda} =&\lambda \int dz d\bar{z}\bigg[\sum_{i}\left(\frac{h_i}{(z-z_i)^2}+\frac{\partial_{z_i}}{z-z_i}\right)\left(\frac{\bar{h}_i}{(\bar{z}-\bar{z}_i)^2}+\frac{\partial_{\bar{z}_i}}{\bar{z}-\bar{z}_i}\right)\notag\\&\qquad  \qquad \qquad\qquad \hspace{3.5cm} \times\langle O(z_1,\bar{z}_1)O(z_2,\bar{z}_2)\rangle_0\bigg]
\end{align}
In the above equation the two point correlation function involving two primary operators for the undeformed CFT is given by, 
\begin{align}
    \langle O_1(z_1,\bar{z}_1)O_2(z_2,\bar{z}_2)\rangle_0=\frac{1}{(z_1-z_2)^{2h}(\bar{z}_1-\bar{z}_2)^{2\bar{h}}}=\frac{1}{z_{12}^{2h}\bar{z}_{12}^{2\bar{h}}}
\end{align}
with $h$ and $\bar{h}$ being the scaling dimensions of the operators. {It is worth emphasizing that within the perturbation framework, the operators are assumed not to be deformed}. Rather one can think that the vacuum state of free CFT got corrections due to the perturbations, resulting in a corrected vacuum expectation value. To examine the Lanczos coefficients and the Krylov complexity  we require the thermal two point function. This can be obtained by the usual conformal map from CFT on a Euclidean plane to a cylinder.\vspace{1em}

\noindent The first order correction to the two point correlation function of two dimensional CFT on a cylinder at a  finite temperature $1/\beta$, due to the $\text{T}\bar{\text{T}}$ deformation is expressed as follows,
\begin{align}\label{TTbarFT2}
 \langle\mathcal{O}\left(w_1, \bar{w}_1\right) \mathcal{O}\left(w_2, \bar{w}_2\right)\rangle_{\beta,\lambda}=&  \langle\mathcal{O}\left(w_1, \bar{w}_1\right) \mathcal{O}\left(w_2, \bar{w}_2\right)\rangle_{\beta,0}\notag\\&\qquad+ \lambda \int d^2 w\left\langle T(w) \bar{T}(\bar{w}) \mathcal{O}\left(w_1, \bar{w}_1\right) \mathcal{O}\left(w_2, \bar{w}_2\right)\right\rangle_{\beta,0}
\end{align}
where the coordinates on cylinder are denoted by $(w, \bar{w})$ and these are related to the coordinates on the complex plane $(z, \bar{z})$ through the following conformal transformation
\begin{equation}\label{coordtrans}
\begin{split}
    &z_i=e^{\frac{2\pi}{\beta}w_i},\quad 
     \bar{z}_i=e^{\frac{2\pi}{\beta}\bar{w}_i};\\&w=x+it, \quad \bar{w}=x-it
\end{split}
\end{equation}
Under the above conformal transformation the  following correlator involving the energy momentum tensors transforms as follows
\begin{align}
&\langle \text{T}(w) \bar{\text{T}}(\bar{w}) \mathcal{O}(w_1, \bar{w}_1) \mathcal{O}(w_2, \bar{w}_2)\rangle_\beta \nonumber\\
= &\prod_{i=1,2}\Big(\frac{2 \pi z_i}{\beta}\Big)^{h_i}\Big(\frac{2 \pi \bar{z}_i}{\beta}\Big)^{h_i}\Big(\frac{2 \pi z}{\beta}\Big)^2\Big(\frac{2 \pi \bar{z}}{\beta}\Big)^2\Big\langle\Big(T(z)-\frac{c}{24 z^2}\Big)\Big(\bar{T}(\bar{z})-\frac{c}{24 \bar{z}^2}\Big) \mathcal{O}(z_1, \bar{z}_1) \mathcal{O}(z_2, \bar{z}_2)\Big\rangle_{0}
\end{align}
Utilizing the above expression in \cref{TTbarFT2} it is easy to obtain,
\begin{align}\label{TTbarFT3}
&\frac{\langle\mathcal{O}(w_1, \bar{w}_1) \mathcal{O}(w_2, \bar{w}_2)\rangle_{\beta,\lambda}}{\langle\mathcal{O}(w_1, \bar{w}_1) \mathcal{O}(w_2, \bar{w}_2)\rangle_{\beta,0}}=  1+  \lambda \Big(\frac{2\pi}{\beta}\Big)^4\int d^2 w |z|^4\frac{\langle(T(z)-\frac{c}{24 z^2})(\bar{T}(\bar{z})-\frac{c}{24 \bar{z}^2}) \mathcal{O}(z_1, \bar{z}_1) \mathcal{O}(z_2, \bar{z}_2)\rangle_{0}}{\langle\mathcal{O}(z_1, \bar{z}_1) \mathcal{O}(z_2, \bar{z}_2)\rangle_0}\nonumber\\
\end{align}
The above equation will be  split  into four different integrals for convenience
\begin{align}
 &\frac{\langle\mathcal{O}(w_1, \bar{w}_1) \mathcal{O}(w_2, \bar{w}_2)\rangle_{\beta,\lambda}}{\langle\mathcal{O}(w_1, \bar{w}_1) \mathcal{O}(w_2, \bar{w}_2)\rangle_{\beta,0}}=  1+  \lambda \Big(\frac{2\pi}{\beta}\Big)^2\Big(I_1+I_2+I_3+I_4\Big)   
\end{align}
where we have set $h=\bar{h}$ for the operator $\mathcal{O}$ and accounted the change in measure of the integral. The four integrals $I_1,I_2,I_3,I_4$ are given as 
\begin{align}
I_1&=|z_{12}|^{4h}\int dz d\bar{z} \, z \bar{z} \langle T(z)\bar{T}(\bar{z}) \mathcal{O}(z_1, \bar{z}_1) \mathcal{O}(z_2, \bar{z}_2)\rangle_0\\
I_2&=-\frac{c |z_{12}|^{4h}}{24 }\int  dz d\bar{z} \, \frac{z}{\bar{z}}\langle T(z) \mathcal{O}(z_1, \bar{z}_1) \mathcal{O}(z_2, \bar{z}_2)\rangle_0\label{I2}\\
I_3&=-\frac{c |z_{12}|^{4h}}{24 }\int  dz d\bar{z} \, \frac{\bar{z}}{z}\langle \bar{T}(\bar{z}) \mathcal{O}(z_1, \bar{z}_1) \mathcal{O}(z_2, \bar{z}_2)\rangle_0\\
I_4&=\left(\frac{c }{24 }\right)^2\int  dz d\bar{z} \, \frac{1}{z\bar{z}}
\end{align}
The integral $I_4$ is divergent and can be shown to depend only on the cutoff. Therefore it will not contribute to any dynamics.  For the other three integrals, we will report the final results upto $\mathcal{O}(\epsilon)$ terms below  and provide  the details of the computation in \cref{app:integrals} 
\begin{align}\label{ITTbar}
    I_1&= 2\pi h^2\biggr[{|z_1|^2 +|z_2|^2\over |z_{12}|^2}\left({4\over \epsilon}+2\gamma-5+2\ln \pi+2\ln|z_{12}|^2+\ln\mu^2\right)\notag\\&\qquad\qquad-\left({2\over \epsilon}+\gamma-3+\ln\pi+\ln|z_{12}|^2+\ln\mu\right)\biggr],\notag\\
    I_2&={c\,h\over 24}\left[2 \pi+\pi \ln|z_1 z_2|^2-{2\pi\over z_{12}}\left(z_1\ln|z_1|^2-z_2\ln|z_2|^2\right)\right],\notag\\
    I_3&={c\,\bar{h}\over 24}\left[2 \pi+\pi \ln|z_1 z_2|^2-{2\pi\over \bar{z}_{12}}\left(\bar{z}_1\ln|z_1|^2-\bar{z}_2\ln|z_2|^2\right)\right].
\end{align}
Finally, utilizing the coordinate transformation given in \cref{coordtrans} in \cref{ITTbar}, we can obtain the two point correlation function on the cylinder. We will compute the autocorrelation function from these results in the next section.\vspace{1em}\\ 
There exist additional families of solvable QFT deformations. For instance, in theories featuring a conserved holomorphic $U(1)$ current, one can define another irrelevant deformation known as the J$\bar{\text{T}}$ deformation \cite{Guica:2017lia, Chakraborty:2018vja, Guica:2021fkv}. Analogous to the T$\bar{\text{T}}$ deformation, the deformed action can be defined as given by the following expression
\begin{equation}
 \frac{\partial S}{\partial \lambda}=\int \sqrt{g}~d^2x~ (\text{J}\bar{\text{T}})_{\lambda}  
\end{equation}
Within the framework of perturbative  analysis, the first order correction to the two point function on the plane from OPEs of the operators are determined as follows \cite{He:2019vzf, Guica:2021fkv},
\begin{align}\label{2pt:planeJT}
  \langle O(z_1,\bar{z}_1)O(z_2,\bar{z}_2)\rangle_{\lambda} &=\notag\\&\lambda \int dz d\bar{z}\left(\frac{q_1}{z-z_1}+\frac{q_2}{z-z_2}\right)\frac{\bar{h}\bar{z}^2_{12}}{(\bar{z}-\bar{z}_1)^2(\bar{z}-\bar{z}_2)^2}\langle O(z_1,\bar{z}_1)O(z_2,\bar{z}_2)\rangle_0
\end{align}
where $q$'s are the charges of the primary operators $O$'s under the operation of the $U(1)$ current. We will write the thermal two point function from \cref{2pt:planeJT} by applying the coordinate transformation given in \cref{coordtrans},
\begin{align}\label{JTbar1}
 \langle\mathcal{O}\left(w_1, \bar{w}_1\right) \mathcal{O}\left(w_2, \bar{w}_2\right)\rangle_{\beta,\lambda}&=  \langle\mathcal{O}\left(w_1, \bar{w}_1\right) \mathcal{O}\left(w_2, \bar{w}_2\right)\rangle_{\beta,0}\notag\\&\quad+  \lambda \int d^2 w\left\langle J(w) \bar{T} (\bar{w}) \mathcal{O}\left(w_1, \bar{w}_1\right) \mathcal{O}\left(w_2, \bar{w}_2\right)\right\rangle_\beta
\end{align}
Considering the transformation properties of J and $\bar{\text{T}}$ operators under \cref{coordtrans},
\begin{align}
&\langle J(w) \bar{T}(\bar{w}) \mathcal{O}(w_1, \bar{w}_1) \mathcal{O}(w_2, \bar{w}_2)\rangle_\beta \notag\\
= &\prod_{i=1,2}\Big(\frac{2 \pi z_i}{\beta}\Big)^{h_i}\Big(\frac{2 \pi \bar{z}_i}{\beta}\Big)^{h_i}\Big(\frac{2 \pi \bar{z}}{\beta}\Big)^2\Big(\frac{2 \pi{z}}{\beta}\Big)\langle J(z)(\bar{T}(\bar{z})-\frac{c}{24 \bar{z}^2}) \mathcal{O}(z_1, \bar{z}_1) \mathcal{O}(z_2, \bar{z}_2)\rangle_{0}
\end{align}
we can manipulate \cref{JTbar1} to yield
\begin{align}\label{JTbarFT4}
\frac{\langle\mathcal{O}(w_1, \bar{w}_1) \mathcal{O}(w_2, \bar{w}_2)\rangle_{\beta,\lambda}}{\langle\mathcal{O}(w_1, \bar{w}_1) \mathcal{O}(w_2, \bar{w}_2)\rangle_{\beta,0}}=&  1+\Big({{2\pi}\over \beta}\Big)  \lambda \int d^2 z\, {\bar{z}}\frac{\langle J(z)(\bar{T}(\bar{z})-\frac{c}{24 \bar{z}^2}) \mathcal{O}(z_1, \bar{z}_1) \mathcal{O}(z_2, \bar{z}_2)\rangle_{0}}{\langle\mathcal{O}(z_1, \bar{z}_1) \mathcal{O}(z_2, \bar{z}_2)\rangle_{0}}\notag\\&=1+\Big({{2\pi}\over \beta}\Big)  \lambda(\mathbb{I}_1+\mathbb{I}_2)
\end{align}
For convenience, we will split the second term on the RHS involving integration into two different integrals with $h=\bar{h}$
\begin{align}
\mathbb{I}_1(z_1,\bar{z}_1;z_2,\bar{z}_2)&=|z_{12}|^{4h}\int dz d\bar{z} \, {\bar{z}} \langle J(z)\bar{T}(\bar{z}) \mathcal{O}(z_1, \bar{z}_1) \mathcal{O}(z_2, \bar{z}_2)\rangle_{0}\\
\mathbb{I}_2(z_1,\bar{z}_1;z_2,\bar{z}_2)&=-\frac{c |z_{12}|^{4h}}{24 }\int  dz d\bar{z} \, {1\over \bar{z}}\langle J(z) \mathcal{O}(z_1, \bar{z}_1) \mathcal{O}(z_2, \bar{z}_2)\rangle_{0}
\end{align}
We provide the detailed computation of these integrals in the appendix \ref{app:jtbar} and the final contributions discarding the ${\cal{O}(\epsilon)}$ terms are,
\begin{equation}
\begin{split}
    \mathbb{I}_1&={\pi\lambda \bar{h} \over 2\bar{z}_{12}}(\bar{z}_1+\bar{z}_2) (q_2-q_1)\left({4\over \epsilon}-5+2\gamma+2\ln\pi+2\ln \mu |z_{12}|^2\right)+{\pi\over \bar{z}_{12}}\lambda\, \bar{h}\, (q_2 \bar{z}_2-q_1 \bar{z}_1)\\\mathbb{I}_2&={\pi c\over 24}\left[\left(q_1+q_2\right)\left({2\over \epsilon}+\gamma+\ln\pi\right)+q_1\ln|z_1|^2+q_2\ln|z_2|^2\right]
\end{split}
\end{equation}
Since the autocorrelation function is related to the unequal time thermal two point function
of the same operator $\mathcal{O}$, the charges $q_1$ and $q_2$ are equal i.e $q_1=q_2=q$. Hence, the results of the above integrals simplify remarkably to,
\begin{equation}\label{JT:intresult}
\begin{split}
    \mathbb{I}_1&={\pi\over \bar{z}_{12}}\lambda\, \bar{h}\, q( \bar{z}_2- \bar{z}_1)\\\mathbb{I}_2&={\pi c\over 24}\left[2q\left({2\over \epsilon}+\gamma+\ln\pi+\ln \mu\right)+q\ln(|z_1 z_2|^2)\right]
\end{split}
\end{equation}
\noindent Analogous to the T$\bar{\text{T}}$ case, we may compute the thermal two point function with deformations expressing \cref{JT:intresult} in cylindrical coordinates and obtain the  corresponding autocorrelation function. Computation of correlation function of the deformed theory non-perturbatively was first proposed in \cite{Cardy:2018sdv} and was also investigated in \cite{Kruthoff:2020hsi, Guica:2021fkv}. These approaches promise to yield highly nontrivial insights into the renormalization flow structure of correlation functions.\vspace{1em}\\
Finally, we consider another marginal deformation to Euclidean CFTs induced by a bilinear operator J$\bar{\text{J}}$ constructed from both the conserved holomorphic and antiholomorphic $U(1)$ currents  \cite{Cardy:2018sdv, Kruthoff:2020hsi, Guica:2021fkv}. From the OPE and the Ward identity, we may obtain the deformed two point function on the complex plane as follows
\begin{align}\label{WardJJB}
\langle J(z) \bar{J}( \bar{z})O(z_1,\bar{z}_1)O(z_2,\bar{z}_2)\rangle_{\lambda} =&\lambda \int dz d\bar{z}\left(\frac{q_1}{z-z_1}+\frac{q_2}{z-z_2}\right)\left(\frac{{\bar{q}}_1}{\bar{z}-\bar{z}_1}+\frac{{\bar{q}}_2}{\bar{z}-\bar{z}_2}\right)\notag\\&\qquad\langle O(z_1,\bar{z}_1)O(z_2,\bar{z}_2)\rangle_0
\end{align}
The deformed thermal correlation function is therefore given by
\begin{align}\label{JJbarFT2}
 \frac{\langle\mathcal{O}\left(w_1, \bar{w}_1\right) \mathcal{O}\left(w_2, \bar{w}_2\right)\rangle_{\beta,\lambda}}{\langle\mathcal{O}\left(w_1, \bar{w}_1\right) \mathcal{O}\left(w_2, \bar{w}_2\right)\rangle_{\beta,0}}=  1+  \lambda \int d^2 w\frac{\left\langle J(w) \bar{J}(\bar{w}) \mathcal{O}\left(w_1, \bar{w}_1\right) \mathcal{O}\left(w_2, \bar{w}_2\right)\right\rangle_\beta}{\langle\mathcal{O}\left(w_1, \bar{w}_1\right) \mathcal{O}\left(w_2, \bar{w}_2\right)\rangle_{\beta,0}}
\end{align}
Utilizing the conformal transformation given in \cref{coordtrans} the required correlator on the cylinder may be related to that on the complex plane as follows
\begin{align}
&\langle J(w) \bar{J}(\bar{w}) \mathcal{O}(w_1, \bar{w}_1) \mathcal{O}(w_2, \bar{w}_2)\rangle_\beta \\
= &\prod_{i=1,2}\Big(\frac{2 \pi z_i}{\beta}\Big)^{h_i}\Big(\frac{2 \pi \bar{z}_i}{\beta}\Big)^{h_i}\Big(\frac{2 \pi \bar{z}}{\beta}\Big)\Big(\frac{2 \pi{z}}{\beta}\Big)\langle J(z)\bar{J}(z) \mathcal{O}(z_1, \bar{z}_1) \mathcal{O}(z_2, \bar{z}_2)\rangle_{0}
\end{align}
Utilizing the above expression and \cref{WardJJB} in \cref{JJbarFT2} reduces the computation of the required thermal two point correlation into evaluation of the following integrals
\begin{align}
    &{\cal Y}_1=\int dz d\bar{z}~\frac{q_1\bar{q}_1}{|z-z_1|^2},\quad{\cal Y}_2=\int dz d\bar{z}~\frac{q_2\bar{q}_2}{|z-z_2|^2}\notag\\&{\cal Y}_3=\int dz d\bar{z}~\frac{q_2}{z-z_2}\frac{\bar{q}_1}{\bar{z}-\bar{z}_1},\quad {\cal Y}_4=\int dz d\bar{z}~\frac{q_1}{z-z_1}\frac{\bar{q}_2}{\bar{z}-\bar{z}_2}
\end{align}
The details of the computation of the above integrals are provided in appendix \ref{app:jjbar}. While ${\cal Y}_1$ and ${\cal Y}_2$ only contribute at  ${\cal O}(\epsilon)$, ${\cal Y}_3$ and ${\cal Y}_4$ contribute at the leading order which leads us to the following result
\begin{align}
\frac{\langle\mathcal{O}\left(w_1, \bar{w}_1\right) \mathcal{O}\left(w_2, \bar{w}_2\right)\rangle_{\beta,\lambda}}{\langle\mathcal{O}\left(w_1, \bar{w}_1\right) \mathcal{O}\left(w_2, \bar{w}_2\right)\rangle_{\beta,0}}=1-\pi\lambda\left(q_1\bar{q}_2+\bar{q}_1{q}_2\right)\left({2\over \epsilon}+\gamma+\ln\pi+\ln \mu|z_{12}|^2\right)+\mathcal{O}(\epsilon)
\end{align}
Once again since we are finally interested in computing the autocorrelation function, we may take $q_1=q_2=q$ to obtain
\begin{align}\label{JJ:intresult}
\frac{\langle\mathcal{O}\left(w_1, \bar{w}_1\right) \mathcal{O}\left(w_2, \bar{w}_2\right)\rangle_{\beta,\lambda}}{\langle\mathcal{O}\left(w_1, \bar{w}_1\right) \mathcal{O}\left(w_2, \bar{w}_2\right)\rangle_{\beta,0}}=1    -2\pi\lambda q^2\left({2\over \epsilon}+\gamma+\ln\pi+\ln \mu|z_{12}|^2\right)+\mathcal{O}(\epsilon)
\end{align}\vspace{1em}\\
One can also construct a general class of integrable deformations of CFTs in particular of the WZW type with non abelian
current-current operators. These are known as $\lambda$-deformations, first constructed in \cite{Sfetsos:2013wia} which reduce to J$\bar{\text{J}}$ deformation at the perturbative limit. In particular the $\lambda $-deformed action reads
\begin{equation}
S_{k,\lambda}(g)=S_{\text{WZW}}(g)+\frac{k}{2\pi} \int d^2z\, J^{a}(z)\,(\lambda^{-1}-D^T)_{ab}\bar J^b(\bar z)
\end{equation}
where $g\in G$ a semi-simple group element, $k$ is the integer level of the undeformed WZW model and $\lambda$ is the coupling constant. Rescaling the current as $J^a\to (1/\sqrt k) J^a$ the perturbative limit reads 
\begin{equation}
S_{k,\lambda}(g)=S_{\text{WZW}}(g)+\frac{\lambda }{2\pi} \int d^2z\, J^{a}(z)\bar J^a(\bar z)
\end{equation}
The deformed two point functions have been computed non-perturbatively in \cite{Georgiou:2016iom, Georgiou:2019jcf} and reads
\begin{equation}\label{lambda}
\langle {\mathcal{O}}(z_1,\bar{ z}_1)\mathcal{O}(z_2,\bar {z}_2)\rangle=\frac{\delta_{IJ}}{z_{12}^{\gamma^{(I)}_{R,R'}(k,\lambda)}\bar{ z}_{12}^{\gamma^{(I)}_{R,'R}(k,\lambda)}}
\end{equation}
Compared to CFTs, here in \cref{lambda} the scaling dimension of operators in the two point function got replaced by  anomalous dimensions $\gamma^{(I)}(k,\lambda)$.\vspace{1em}

\noindent In the next section, we proceed to evaluate the Lanczos coefficients and the  Krylov complexity for perturbative T$\bar{\text{T}}$, J$\bar{\text{T}}$ and J$\bar{\text{J}}$ deformations. For the case of T$\bar{\text{T}}$, the Lanczos coefficients and Krylov exponent extracted from the large time behaviour of Krylov complexity receive correction at the first order in $\lambda$. We demonstrate that the corrected Lanczos coefficients do not exhibit linear growth with $n$ in the valid perturbation regime. Moreover, through numerical analysis, we illustrate that the Krylov complexity {for positive value of deformation parameter} seemingly contravenes the conjectured bound for the Krylov exponent within the finite dimensional Krylov space, proposed in \cite{Parker:2018yvk}. In contrast to this
for the J$\bar{\text{T}}$ case, the Lanczos coefficients and  the Krylov complexity undergo no corrections at the first order in $\lambda$. On the other hand, we will also demonstrate that although  the Lanczos coefficients and the Krylov complexity receive corrections for  the J$\bar{\text{J}}$ deformation at first order in $\lambda$, both the large-$n$ behaviour of $b_n$ and the Krylov exponent are exactly same as that of the undeformed CFT. We do not expect any corrections for both the Lanczos coefficients and Krylov complexity due to the deformation in \cref{lambda}, since there is no nontrivial time dependent corrections involved in the two point function compared to the undeformed CFT case.

\section{Krylov complexity in deformed CFTs}\label{Sec:main}

In this section, we very briefly review the notion of Krylov complexity and Lanczos coefficients\footnote{For detailed reviews we refer the reader to \cite{Parker:2018yvk, Avdoshkin:2019trj, Dymarsky:2019elm, Rabinovici:2020ryf, Caputa:2022eye, Caputa:2021sib, Balasubramanian:2021mxo, Balasubramanian:2022tpr, Rabinovici:2021qqt, Dymarsky:2021bjq, Rabinovici:2022beu, Fan:2022xaa}} and then proceed to compute these quantities for perturbative T$\bar{\text{T}}$, J$\bar{\text{T}}$ and J$\bar{\text{J}}$ deformations of a two dimensional Euclidean conformal field theory which were described in the previous section.

\subsection{Review of Lanczos coefficients and  Krylov complexity}
The evolution of an operator in the Heisenberg picture is given by
\begin{equation}
	\mathcal{O}(t)=e^{i H t} \mathcal{O}(0) e^{-i H t}
\end{equation}
Upon expanding the above expression order by order in $t$ leads to nested commutators of the operator of interest at $t=0$, whose span forms an invariant subspace known as the Krylov space, $\operatorname{span}\left\{\mathcal{L}^{n} \mathcal{O}\right\}_{n=0}^{+\infty}=\operatorname{span}\{\mathcal{O},[H, \mathcal{O}],[H,[H, \mathcal{O}]],\ldots\}$. The ${\mathcal{L}}$ itself is known as the Liouvillian whose action on operators is defined by the commutator $[H,.]$. An orthonormal basis for the Krylov space mentioned above can be obtained utilizing the  Gram-Schmidt procedure, which in this particular case is known as the Lanczos algorithm \cite{Viswanath:2008, Parker:2018yvk}. This algorithm begins with the initial operator as the zeroth element i.e $\mid \mathcal{O}(t=0)):=\mid \mathcal{O}_{0})$ and leads to the entire basis given as follows
\begin{align}
\mid \tilde{\mathcal{O}}_{n})&=\mathcal{L} \mid \mathcal{O}_{n-1})-b_{n-1} \mid \mathcal{O}_{n-2})-a_{n-1}|{\cal O}_{n-1})\label{LactO}\\
\text{with}~~\mid \mathcal{O}_{n})&=b_{n}^{-1} \mid \tilde{\mathcal{O}}_{n}), \quad b_{n}=( \tilde{\mathcal{O}}_{n} \mid \tilde{\mathcal{O}}_{n})^{1 / 2}, \quad a_n=(\mathcal{O}_{n}\mid \mathcal{L}\mid \mathcal{O}_{n})
\label{Basis2}
\end{align}
The coefficients $b_n$ and $a_n$ are called the Lanczos coefficients. 
The construction of the Krylov space, requires a positive semi-definite inner product for instance like Wightman inner product 
 \begin{align}\label{2pt}
(	\mathcal{O}_1 \mid \mathcal{O}_2)=\operatorname{Tr}\left(\rho^{\frac{1}{2}} \mathcal{O}_1 ^{\dagger} \rho^{\frac{1}{2}} \mathcal{O}_2\right)
\end{align}
where $\rho=e^{-\beta H}$ is the thermal density matrix. Therefore, the operator at any time can be expanded in the Krylov basis
\begin{align}\label{Ot}
	| \mathcal{O}(t))=\sum_{n} i^{n} \varphi_{n}(t) | \mathcal{O}_{n})
\end{align}
In the above expression $\varphi_{n}(t)$ are treated as the probability amplitude for the operator to be in the basis state $| \mathcal{O}_{n})$ and the Heisenberg equation for the operator reduces to a one dimensional recursive equation resembling that of an electron hopping on a 1d chain which is as follows
\begin{align}\label{KWFEM}
	\partial_{t} \varphi_{n}(t)=b_{n} \varphi_{n-1}(t)-b_{n+1} \varphi_{n+1}(t)+i a_n\varphi_{n}(t)
\end{align}
This recursion relation has to be solved with the boundary conditions  $\varphi_{-1}(t)=0$, along with $b_0=0$, and $\varphi_n(0)=\delta_{n,0}$. The Krylov complexity characterizes the average position of $\phi_n$'s in the 1d chain and therefore can be defined as
\begin{align}\label{Kdef}
K_{\mathcal{O}}(t) =\left(\mathcal{O}(t)|n|\mathcal{O}(t)\right)=\sum_{n=0}^{\infty} n\left|\varphi_{n}(t)\right|^{2}
\end{align}

Having reviewed the notion of the Lanczos coefficients and the Krylov complexity of an operator of a quantum system, we now proceed towards their computation in a two dimensional CFT whose Hamiltonian is deformed by T$\bar{\text{T}}$, J$\bar{\text{T}}$ and J$\bar{\text{J}}$ composite operators. We will assume that the primary operators of the seed CFT are undeformed and the time evolution of these operators depend on the Hamiltonian of the theory which is changing along the flow. Thus finite time operators
will change due to their dependence on deformed Hamiltonian \cite{Kruthoff:2020hsi},
\begin{align}
 \mathcal{O}(t,0)= e^{itH_{\lambda}} \mathcal{O}(0,0) e^{-itH_{\lambda}}
\end{align}
where $H_{\lambda}=\int dx T_{00}^{(\lambda)}(0,x)$. Note that the zeroth order wave function $\varphi_0(t)$ which is the primary quantity crucial for all the computations is known as the autocorrelation function. For a Hermitian operator $\mathcal{O}$, this  autocorrelation function is given by
\begin{align}\label{auto}
    C_0(t) \equiv\varphi_{0}(t)&=(\mathcal{O}(t)|\mathcal{O}(0))\nonumber\\
    &=\langle \mathcal{O}(t-i\beta/2)\mathcal{O}(0)\rangle_{\beta}
\end{align}
where the last line which relates the autocorrelation function to the thermal two point function is obtained by utilizing the definition of Wightman inner product in \cref{2pt}, which incorporates a shift in Euclidean time by $\beta/2$ to avoid the divergence. From the above expression one can observe that the autocorrelation function is directly related to the thermal two point function. One can compute moments from the Taylor series expansion coefficients of $C_0(t)$ around $t=0$. These moments satisfy a recursion relation which is ultimately utilized to compute the Lanczos coefficients $b_n$. In \cite{Parker:2018yvk} it was conjectured that $b_n$ satisfy a linear growth for large $n$, for maximal chaotic systems evolving under a Hamiltonian involving local interactions. This aspect can also be understood from the high frequency behaviour of the power spectrum obtained from the Fourier transform of the autocorrelation function,
\begin{equation}\label{power}
    f^2(\omega)={1\over 2\pi}\int_{-\infty}^{\infty}~dt e^{i\omega t}C_0(t)
\end{equation}
If $C_0(t)$ has a pole at $t=i/\omega_0$, then $f^2(\omega)$ has an exponential asymptotic behaviour
\begin{equation}\label{tail}
    f^2(\omega)|_{\omega\rightarrow \infty}\sim  e^{-\omega/\omega_0}
\end{equation}
It is important to note that asymptotic linear growth of $b_n$ implies exponential decay of high frequency tail of the power spectrum but not vice-versa \cite{Avdoshkin:2022xuw}.
\vspace{1em}

\noindent In the previous section,  we have reviewed the thermal two point functions involving primary operators of a two dimensional CFT with the first order perturbative correction induced by T$\bar{\text{T}}$, J$\bar{\text{T}}$ and J$\bar{\text{J}}$ deformations following \cite{He:2019vzf,Kraus:2018xrn}. We will utilize these results and translate them in Euclidean time by $\beta/2$ to obtain the {series expansion}
\begin{align}
C_0^\lambda(t)&=\langle \mathcal{O}(t-i\beta/2)\mathcal{O}(0)\rangle_{\beta,\lambda}\nonumber\\
&=\langle \mathcal{O}(t-i\beta/2)\mathcal{O}(0)\rangle_{\beta,0}+\lambda \langle \mathcal{O}(t-i\beta/2)\mathcal{O}(0)\rangle^{\lambda}_{\beta}+O(\lambda^2)
\end{align}
where $\langle \mathcal{O}(t-i\beta/2)\mathcal{O}(0)\rangle^{\lambda}_{\beta}$ denotes the first order correction to the autocorrelation function. One should note that this autocorrelation function needs to be properly normalised with a factor of $C_0^\lambda(0)$ to obtain the autocorrelation function $$C_0(t)={C_0^\lambda(t)\over C_0^\lambda(0)},$$ before proceeding with the computations for Lanczos coefficients.
Determining the required correlators involve evaluating  numerous integrals whose computational details are in appendix \ref{app:integrals}. Once the autocorrelation function is determined, the Lanczos coefficients are computed utilizing either the moment method or the Toda chain technique. Since these two methods are explained in quite extensive details in several recent literature, we do not discuss them here, however we refer the interested readers to detailed discussions in \cite{Parker:2018yvk,Viswanath:2008,Dymarsky:2019elm,Kundu:2023hbk}. From the autocorrelation function and the Lanczos coefficients the operator wave functions are then obtained by solving the recursion relation in \cref{KWFEM}. Finally, the Krylov complexity is determined using \cref{Kdef}.

\subsection{T$\bar{\text{T}}$}
 In this section we will focus on the final results and discussion of the plausible physical interpretations of the behaviour of Krylov complexity under the above described deformation.
\subsubsection{Autocorrelation function and  Lanczos coefficients}
\begin{figure}[h!]
    \centering
    \begin{subfigure}[h]{0.5\textwidth}
        \centering
        \includegraphics[height=1.75in]{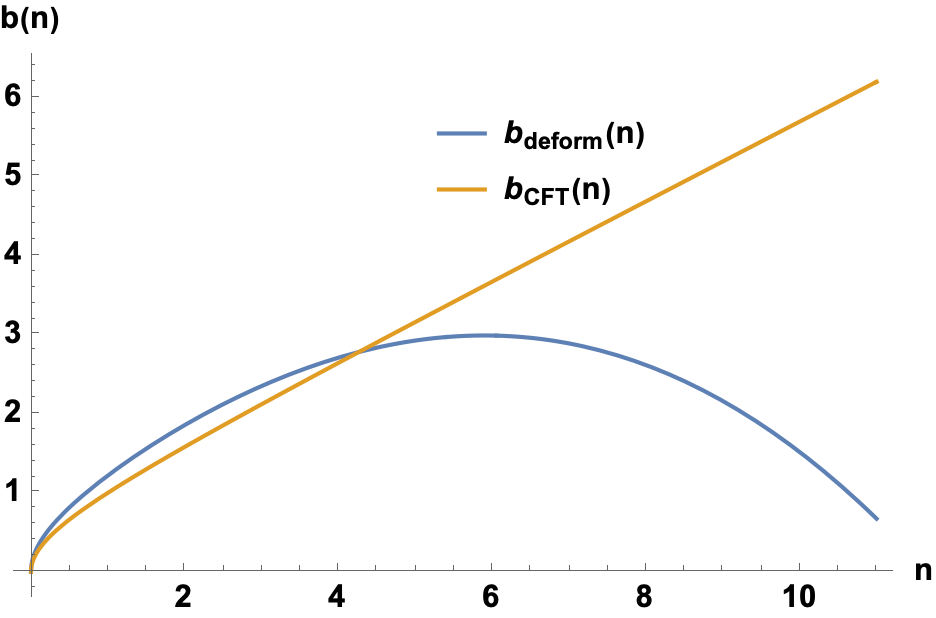}
        \caption{$\lambda=0.07$}
    \end{subfigure}%
    \hfill
    \begin{subfigure}[h]{0.5\textwidth}
        \centering
        \includegraphics[height=1.75in]{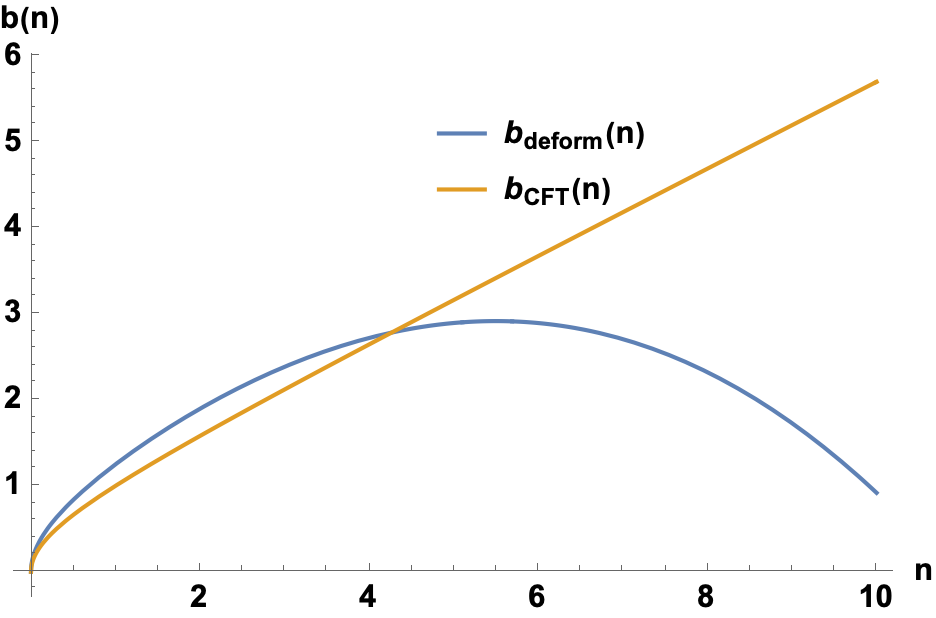}
        \caption{$\lambda=0.08$}
    \end{subfigure}
    \hfill 
    \begin{subfigure}[h]{0.5\textwidth}
        \centering
        \includegraphics[height=1.75in]{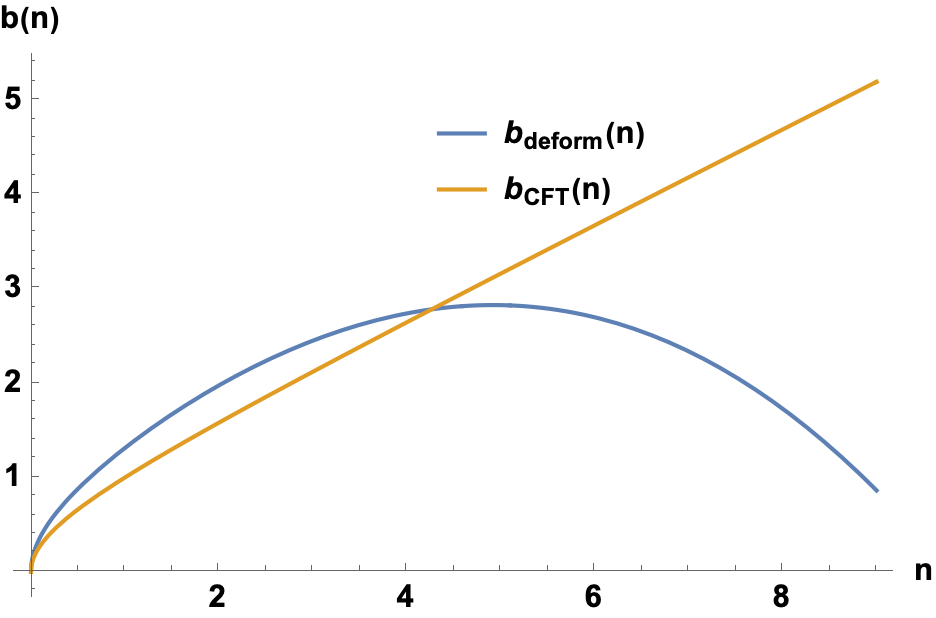}
        \caption{$\lambda=0.10$}
    \end{subfigure}%
     \hfill 
    \begin{subfigure}[h]{0.5\textwidth}
        \centering
        \includegraphics[height=1.75in]{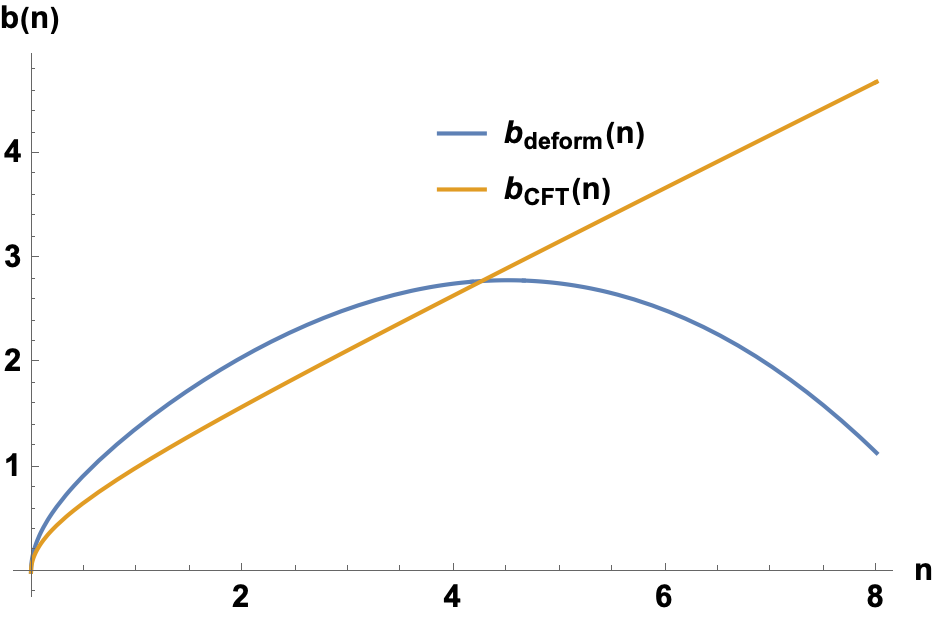}
        \caption{$\lambda=0.12$}
    \end{subfigure}
     \hfill
    \caption{Lanczos coefficients of a two dimensional CFT contrasted with the same post T$\bar{\text{T}}$ deformation until the perturbation breaks down. $b_n$ for the deformed theory will be greater than the undeformed one until $n=4$. A quick numerical calculation will show from \cref{eq:btt} that $b_n=b_0(n)$, when $n=4.265$.}
    \label{fig:TTbLanczosfig}
\end{figure}

We compute the autocorrelation function for a T$\bar{\rm T}$ deformed 2D CFT from \cref{ITTbar}, which yields
\begin{align}\label{autocorrelationTT}
C_0^\lambda(t)=&{1\over \left(\cosh{\pi t\over \beta}\right)^{2\Delta}}+\tilde{\lambda} \bigg\{f_1(\nu_1,\beta,t)-f_2(\nu_2,\beta,t)\bigg\}
\end{align}
with
\begin{align}
f_1(\nu_1,\beta,t)&={\pi\Delta^2\over \left(\cosh{\pi t\over \beta}\right)^{2\Delta+2}}\left[\nu_1+\frac{1}{2}\ln \left(\cosh {\pi t\over \beta}\right)^2\right]\\
   f_2(\nu_2,\beta,t)&= {\pi\Delta^2\over  \left(\cosh{\pi t\over \beta}\right)^{2\Delta}}\left[\nu_2+\frac{1}{2}\ln \left(\cosh {\pi t\over \beta}\right)^2\right]
\end{align}
where we have denoted
\begin{align}
&\tilde{\lambda}=\Big(\frac{2\pi}{\beta}\Big)^2 \lambda,\qquad h=\bar{h}=\Delta/2\notag\\
  &\nu_1={1\over 4}\bigg[{4\over \epsilon}+2\gamma-5+2\ln \pi\mu+4\ln 4\bigg],\notag\\&\nu_2={1\over 2}\bigg[{2\over \epsilon}+\gamma-3+\ln \pi\mu+2\ln 4\bigg]=\nu_1-\frac{1}{4}.
\end{align}
Here $\mu$ is an arbitrary mass scale appearing due to the dimensional regularization procedure. We then calculate the first order corrections to the Lanczos coefficients, given by 
\begin{align}\label{eq:btt}
b_n=b_0(n)+\tilde{\lambda} b_1(n)\bigg\{ \nu_1 - b_2(n) \bigg\}+{\cal O}(\lambda^2),  
\end{align}
where,
\begin{align}
    b_0(n)&=\frac{\pi}{\beta }\sqrt{n (2 \Delta +n-1)}\\
     b_1(n)&=\frac{\pi ^2 \Delta  \sqrt{n (2 \Delta +n-1)} (2 \Delta +2 n-1)}{2 \beta  (2 \Delta +1)}=b_0(n) {\pi\Delta(2 \Delta +2 n-1)\over(2 \Delta +1)}\\
      b_2(n)&=\frac{4 \Delta  (n-1)}{(2 \Delta+1) (2 \Delta +2 n-1)}-\frac{1}{2\Delta }+ H_{n+2 \Delta -1}- H_{2 \Delta -1}
\end{align}
where $H_n=\sum_{k=1}^n{1\over k}$ and $b_0(n)$ represents the undeformed case \cite{Dymarsky:2021bjq}. Since the operator growth hypothesis is concerned with the asymptotic large-n behaviour of Lanczos coefficients, let us now examine the same utilizing the expressions above. One can easily check that the $b_0(n)$ grow linearly whereas the $O(\lambda)$ contributions involving $b_1(n)$ and $b_1(n)b_2(n)$ both grows quadratically. However since  they are multiplied by a perturbative parameter $\lambda$, their contributions are small for low values of $n$. Furthermore, these contributions turn out to be negative  for $n>4$ with positive value of $\lambda$, resulting in a quick dampening of the growth as depicted in the plots given in Fig. \ref{fig:TTbLanczosfig}. \vspace{1em}

\subsubsection{ Krylov complexity}
Utilizing the Lanczos coefficients determined above, the operator wave functions  are obtained by solving the recursive one dimensional equation given by \cref{KWFEM} perturbatively. We also set $\Delta=2$ for computational convenience (the behaviour of complexity is similar for other values of $\Delta$ as well). The probabilities corresponding to the operator wave functions are as follows
\begin{align}
|\phi_n(t)|^2=P_0(n,t)-\tilde{\lambda}\{P_1(n,t)+\nu_1P_2(n,t)\}+{\cal O}(\lambda^2)
\end{align}
where $P_0(n,t)$ is the unperturbed probability and $P_1,P_2$ are the corrections due to the deformation. The functional form of $P_0,P_1,P_2$  are given as follows
\begin{align}
P_0(n,t)&=\frac{1}{6} (n+1) (n+2) (n+3)~ \text{sech}^8\left(\frac{t}{2}\right) \tanh ^{2 n}\left(\frac{t}{2}\right)\nonumber\\
P_1(n,t)&=	\pi f_0^2(n,t)\sum_{i=1}^{12} g_i(n,t)\nonumber\\
P_2(n,t)&=\pi f_0^2(n,t)\sum_{i=1}^{5} h_i(n,t)
\end{align}
We have provided all the functions $f_0,g_i,h_i$ in \cref{OPAT}.\vspace{1em}

\noindent The Krylov complexity itself is not analytically tractable and hence we resort to numerical computations. Also note that since $b_n$'s go to zero after a reasonably large-$n$, we should not be summing the above probabilities all the way upto infinity to obtain the Krylov complexity. In addition, it is pertinent to inquire about the reliability of the Krylov complexity derived from a perturbative autocorrelation function, questioning until which point in time it remains valid.

\begin{figure}[h!] 
    \centering
    \begin{subfigure}[h]{0.5\textwidth}
        \centering
        \includegraphics[height=1.75in]{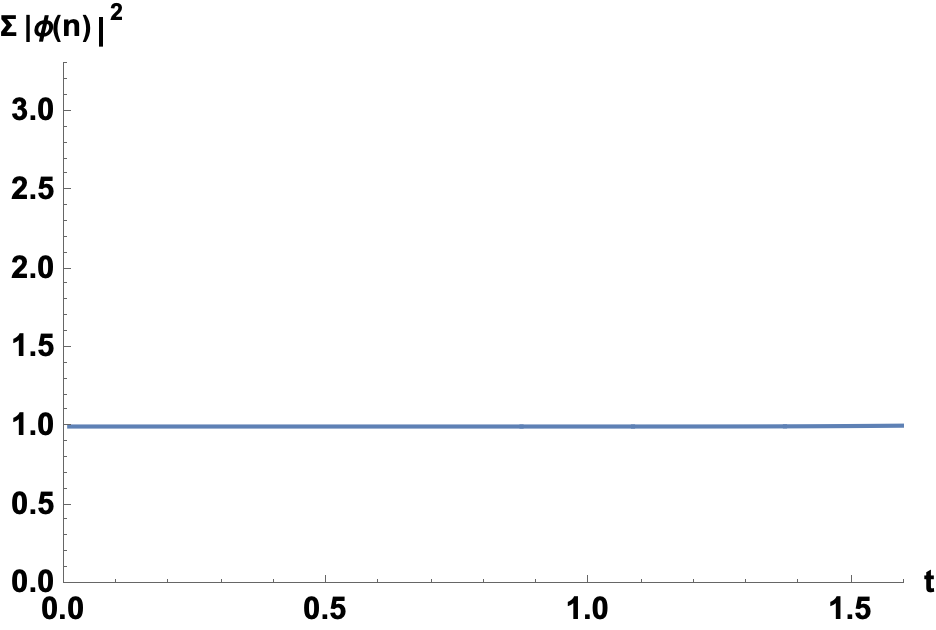}
        \caption{$\lambda=0.07$}
    \end{subfigure}%
    \hfill
    \begin{subfigure}[h]{0.5\textwidth}
        \centering
        \includegraphics[height=1.75in]{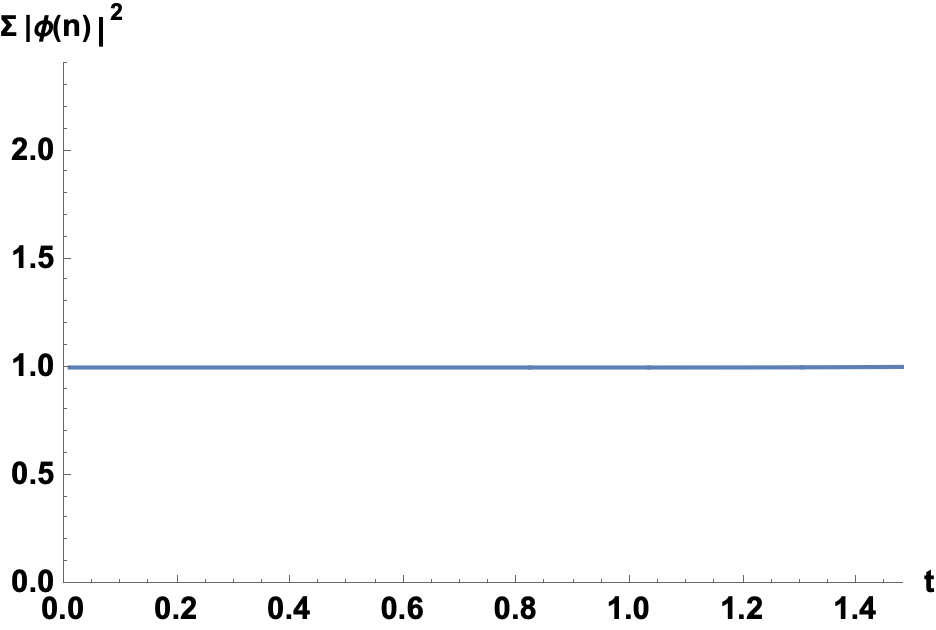}
        \caption{$\lambda=0.08$}
    \end{subfigure}
    \hfill 
    \begin{subfigure}[h]{0.5\textwidth}
        \centering
        \includegraphics[height=1.75in]{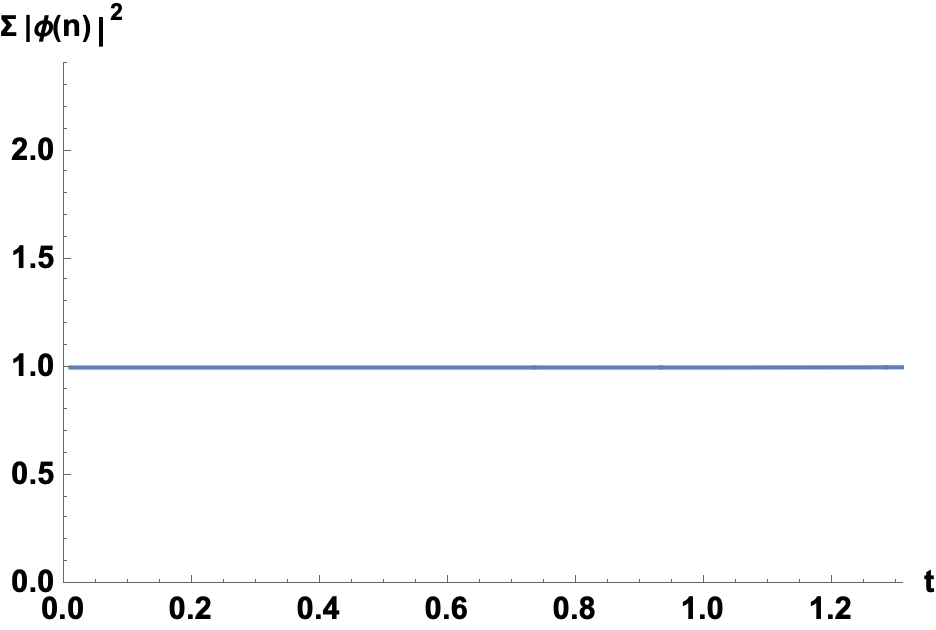}
        \caption{$\lambda=0.10$}
    \end{subfigure}%
     \hfill 
    \begin{subfigure}[h]{0.5\textwidth}
        \centering
        \includegraphics[height=1.75in]{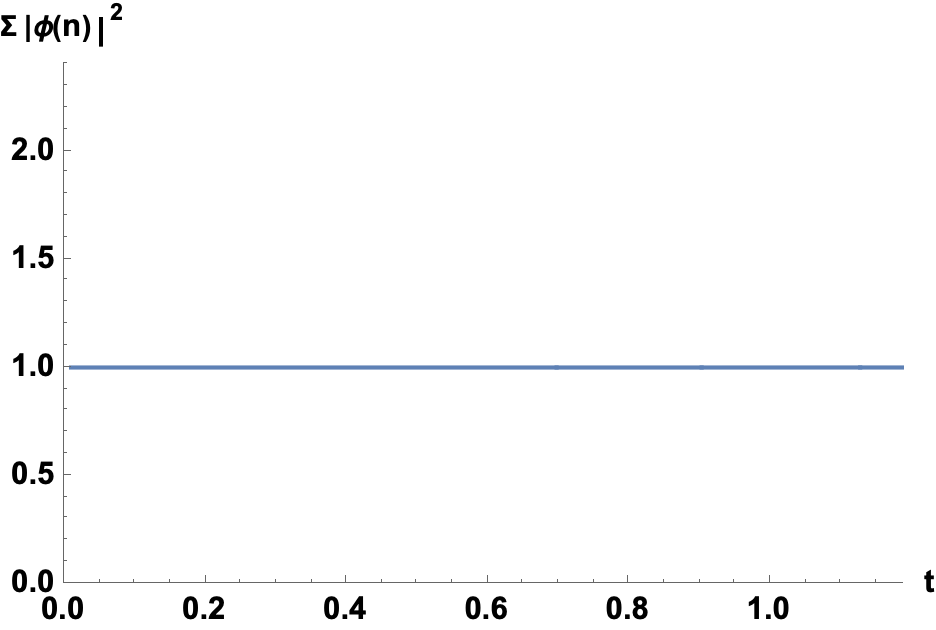}
        \caption{$\lambda=0.12$}
    \end{subfigure}
     \hfill
   \caption{Sum of absolute squares of Probability   post $T\bar{T}$ deformation until the perturbation for the autocorrelation function breaks down for various values of the deformation parameter $\lambda$.}
    \label{probttbar}
\end{figure}
\begin{figure}[h!] 
    \centering
    \begin{subfigure}[h]{0.5\textwidth}
        \centering
        \includegraphics[height=1.75in]{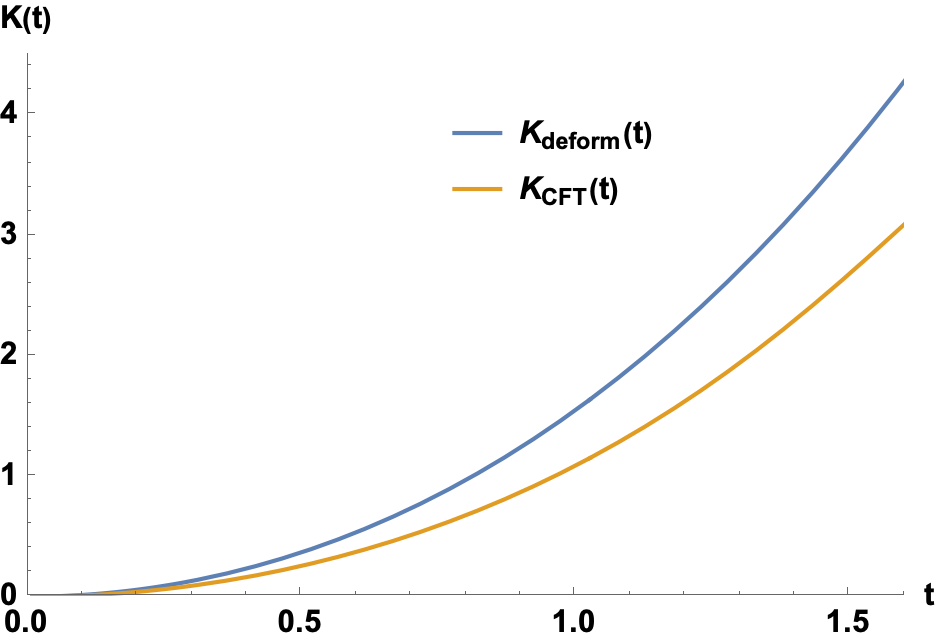}
        \caption{$\lambda=0.07$}
    \end{subfigure}%
    \hfill
    \begin{subfigure}[h]{0.5\textwidth}
        \centering
        \includegraphics[height=1.75in]{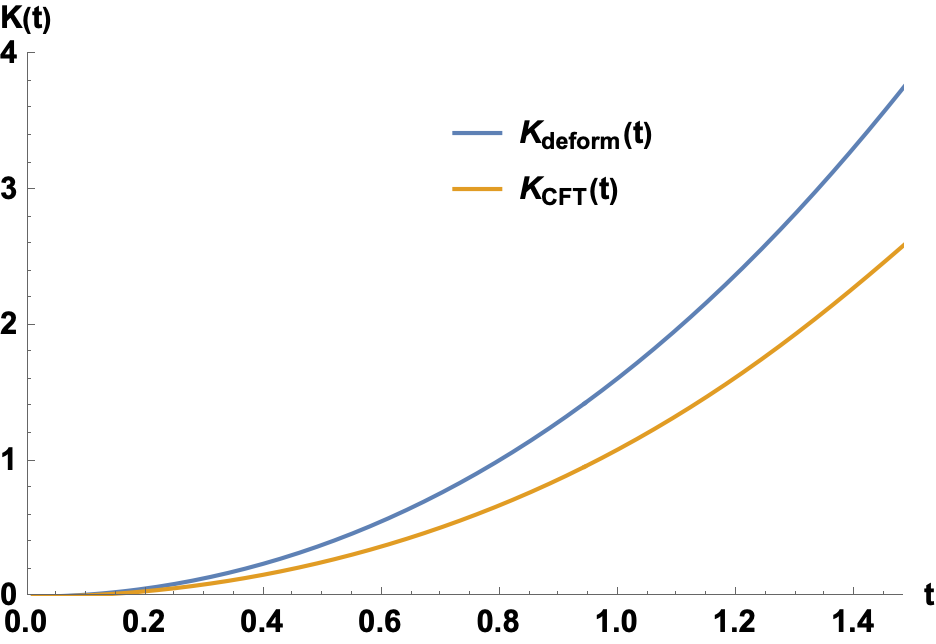}
        \caption{$\lambda=0.08$}
    \end{subfigure}
    \hfill 
    \begin{subfigure}[h]{0.5\textwidth}
        \centering
        \includegraphics[height=1.75in]{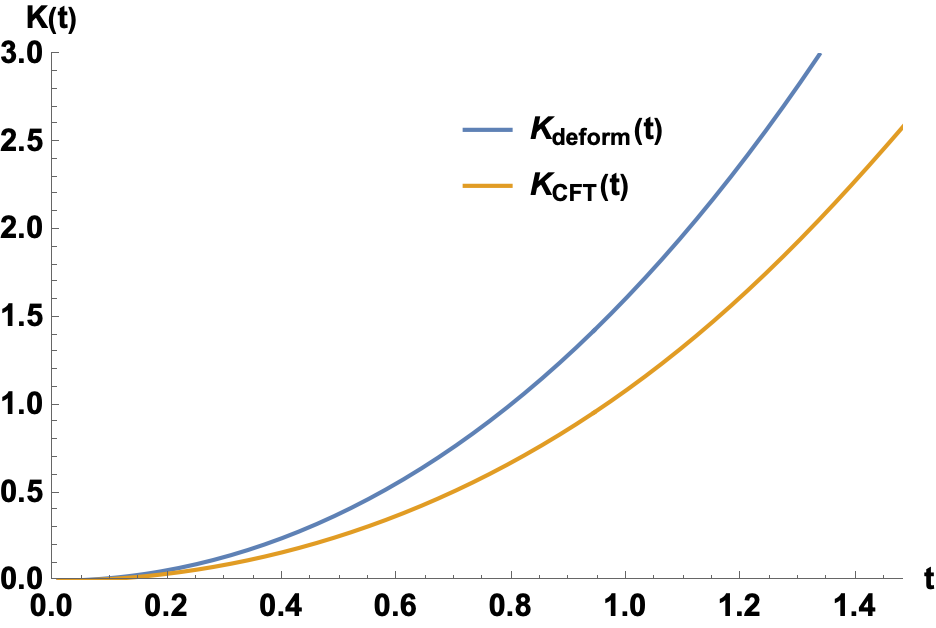}
        \caption{$\lambda=0.10$}
    \end{subfigure}%
     \hfill 
    \begin{subfigure}[h]{0.5\textwidth}
        \centering
        \includegraphics[height=1.75in]{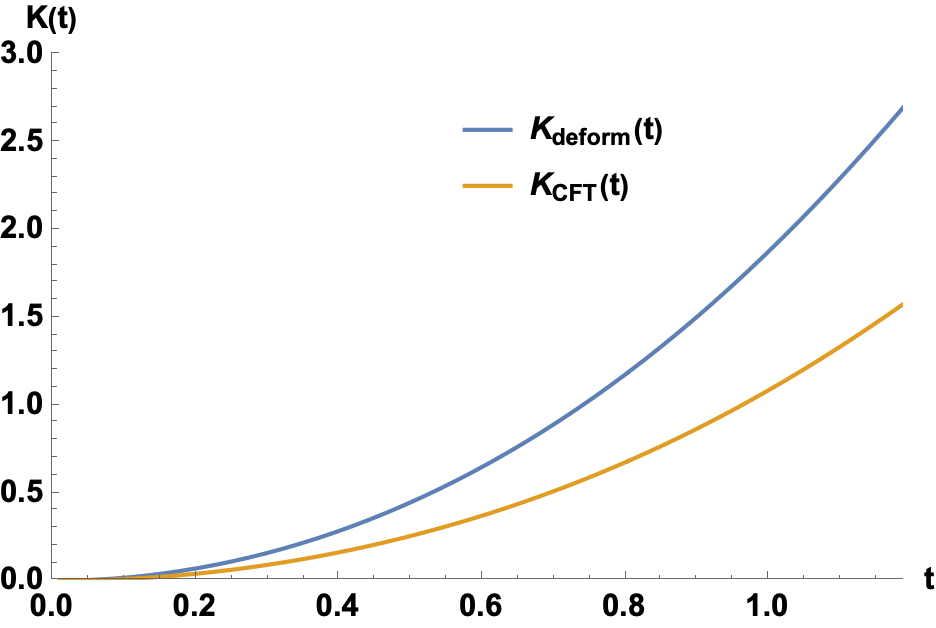}
        \caption{$\lambda=0.12$}
    \end{subfigure}
     \hfill
   \caption{Krylov complexity of a two dimensional CFT contrasted with the the same post T$\bar{\rm T}$  deformation for various values of the deformation parameter $\lambda$. }
    \label{KCttbar}
\end{figure}

\noindent This question was also addressed in \cite{Bhattacharjee:2022ave} where the behaviour of Krylov complexity was examined numerically from a perturbative autocorrelation function. As described in \cite{Bhattacharjee:2022ave}, there are two time scales involved in such an analysis. There is a $t_{\rm max}$ which characterizes the time at which the perturbative series corresponding to the autocorrelation function breaks down i.e terms of $O(\lambda^0)$ and $O(\lambda)$ in the autocorrelation function start competing with each other. However, there is another time scale $t_c$ which emerges as a consequence of truncating the infinite series in \cref{Kdef} to a finite sum, beyond which the probabilities' sum $\sum_i |\phi_n(t)|^2$ begins to decay to values less than 1. The numerics will be only valid if $t_c>t_{\rm max}$ otherwise the finite sum effects start distorting the behaviour of the Krylov complexity before the perturbation breaks down. Taking all these into account by choosing appropriate $\lambda$ such that $t_c>t_{\rm max}$ the behaviour of Krylov complexity is depicted in the plot given in Fig \ref{KCttbar}, while the probability conservation is shown in Fig. \ref{probttbar}.\vspace{1em}

\noindent Note that the plot in Fig \ref{KCttbar} seems to indicate that the Krylov complexity is growing faster compared to the undeformed CFT one with an exponent $\lambda_K>\frac{2\pi}{\beta}$. Several comments are in place here as this seems to be violating the bound conjectured in \cite{Parker:2018yvk}. Note that this is a perturbative result and it is possible that the apparent violation might go away once higher order corrections are taken into account \cite{Avdoshkin:2022xuw}. One analogous example is the case of higher spin symmetry where a violation of the MSS bound for the Lyapunov exponent obtained from OTOCs was observed if one takes into account a finite number of higher spin conformal blocks \cite{Perlmutter:2016pkf}. However, it was argued that if one takes into account all the higher spin blocks (an infinite number of them) they may arrange into a geometric series which can be summed and the resultant Lyapunov exponent shows no violation of the bound anymore.\vspace{1em}

\noindent For a comprehensive understanding we will also consider negative value of deformation parameter. So far we were following the convention of \cite{McGough:2016lol, Kraus:2018xrn} with the deformation parameter $\lambda>0$, which is canonically known as the "bad sign". The origin of the name came from the fact that at sufficiently large energy of the undeformed theory the deformed spectrum become complex. The advantage of this sign is the observed holographic duality between the T$\bar{\rm T}$ deformed theory and putting Dirichlet boundary condition at some finite radius of AdS space. On the other hand one can also consider deformations with $\lambda<0$ corresponding to the "good sign", where the undeformed energy can be taken to be infinitely large. Interestingly in this case the T$\bar{\rm T}$ deformed CFT shares intriguing similarities with a class of non-local theories called little string theory which is in turn dual to gravity theories on linear dilaton background \cite{Giveon:2017nie}.

\begin{figure}[h!]
    \centering
    \begin{subfigure}[h]{0.5\textwidth}
        \centering
        \includegraphics[height=1.75in]{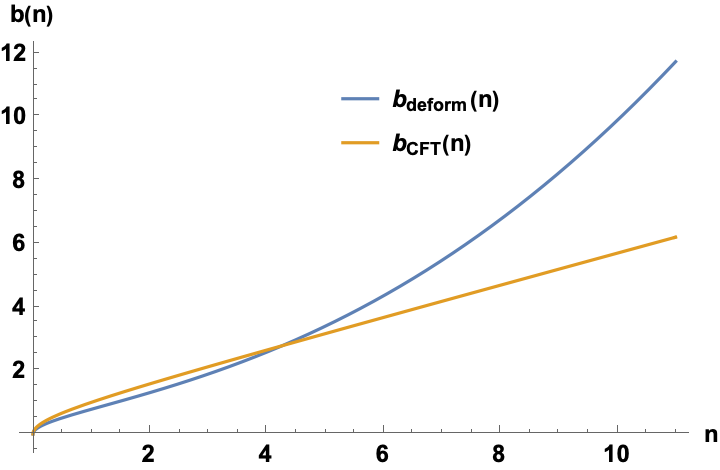}
        \caption{$\lambda=-0.07$}
    \end{subfigure}%
    \hfill
    \begin{subfigure}[h]{0.5\textwidth}
        \centering
        \includegraphics[height=1.75in]{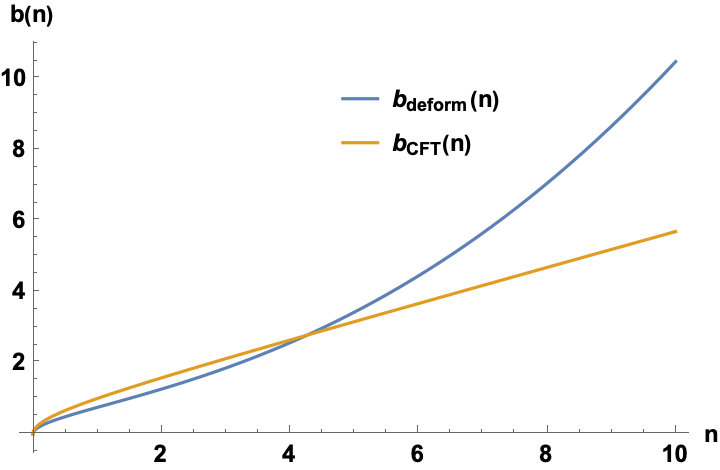}
        \caption{$\lambda=-0.08$}
    \end{subfigure}
    \hfill 
    \begin{subfigure}[h]{0.5\textwidth}
        \centering
        \includegraphics[height=1.75in]{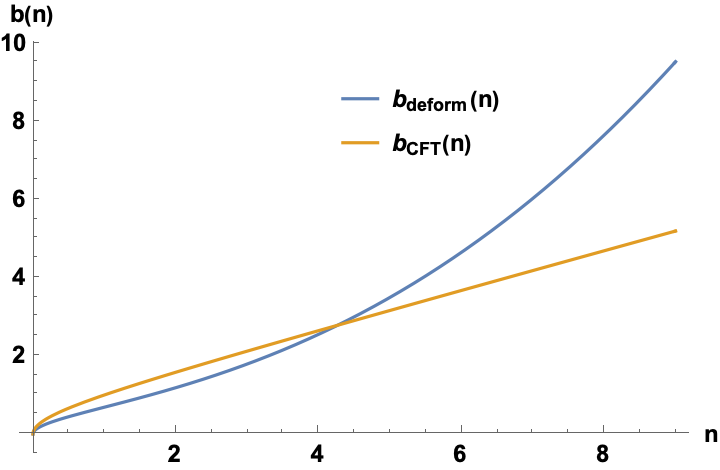}
        \caption{$\lambda=-0.10$}
    \end{subfigure}%
     \hfill 
    \begin{subfigure}[h]{0.5\textwidth}
        \centering
        \includegraphics[height=1.75in]{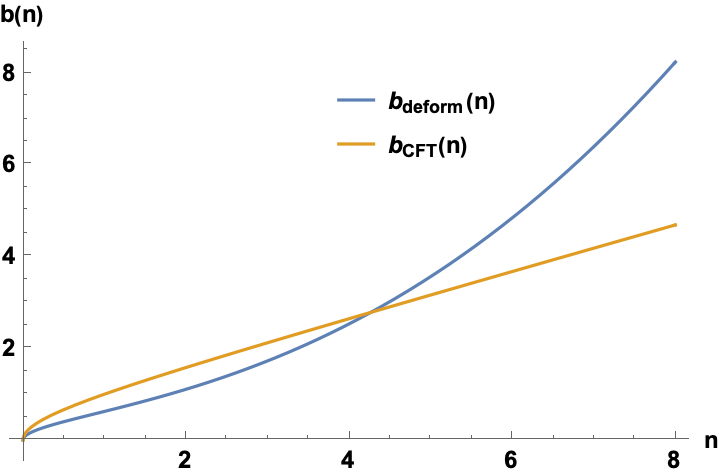}
        \caption{$\lambda=-0.12$}
    \end{subfigure}
     \hfill
    \caption{Lanczos coefficients of a two dimensional CFT contrasted with the same post T$\bar{\text{T}}$ deformation until the perturbation breaks down for negative values of the deformation parameter.}
    \label{fig:TTbLanczosfig2}
\end{figure}

\begin{figure}[h!] 
    \centering
    \begin{subfigure}[h]{0.5\textwidth}
        \centering
        \includegraphics[height=1.6in]{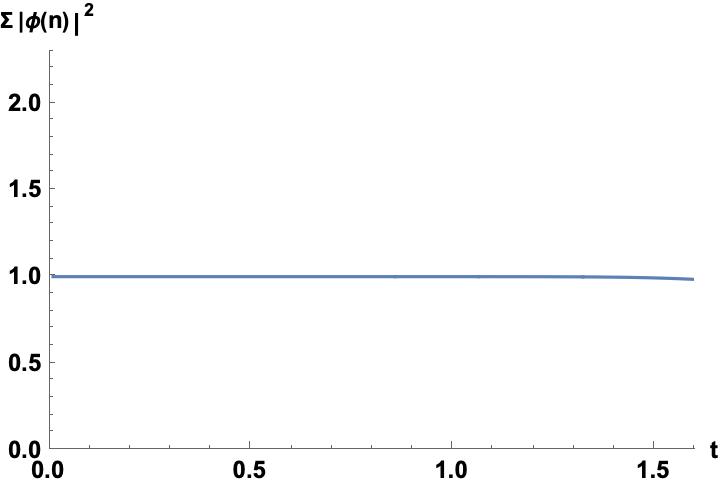}
        \caption{$\lambda=-0.07$}
    \end{subfigure}%
    \hfill
    \begin{subfigure}[h]{0.5\textwidth}
        \centering
        \includegraphics[height=1.6in]{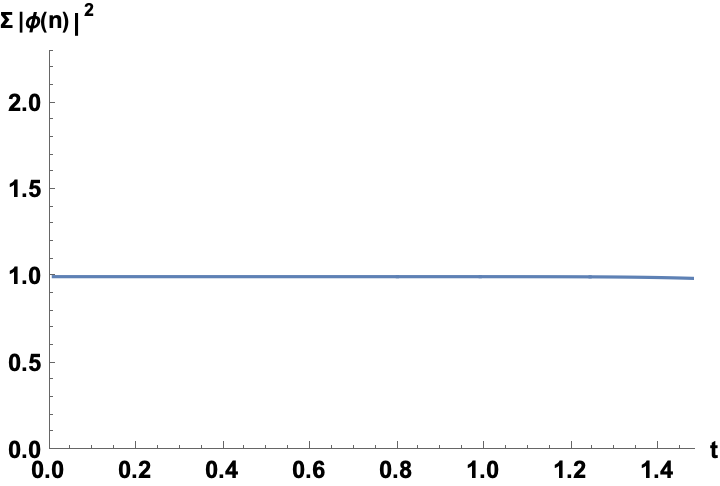}
        \caption{$\lambda=-0.08$}
    \end{subfigure}
    \hfill 
    \begin{subfigure}[h]{0.5\textwidth}
        \centering
        \includegraphics[height=1.6in]{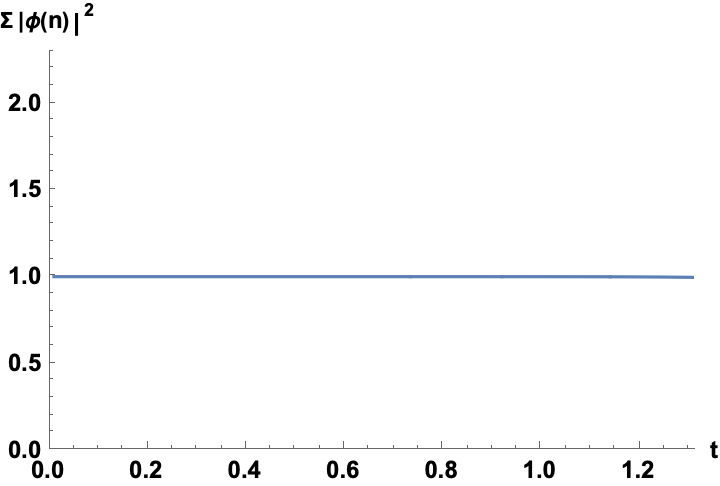}
        \caption{$\lambda=-0.10$}
    \end{subfigure}%
     \hfill 
    \begin{subfigure}[h]{0.5\textwidth}
        \centering
        \includegraphics[height=1.6in]{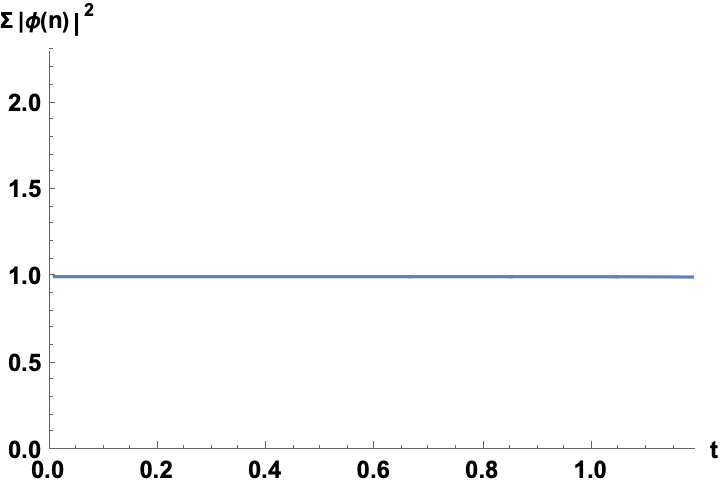}
        \caption{$\lambda=-0.12$}
    \end{subfigure}
     \hfill
   \caption{Sum of absolute squares of Probability   post $T\bar{T}$ deformation until the perturbation for the autocorrelation function breaks down for  negative values of the deformation parameter $\lambda$.}
    \label{probttbars}
\end{figure}

 \begin{figure}[h!] 
    \centering
    \begin{subfigure}[h]{0.5\textwidth}
        \centering
        \includegraphics[height=1.75in]{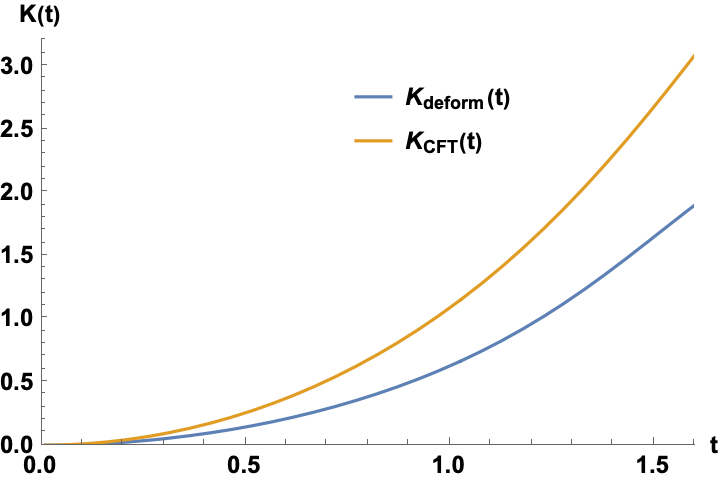}
        \caption{$\lambda=-0.07$}
    \end{subfigure}%
    \hfill
    \begin{subfigure}[h]{0.5\textwidth}
        \centering
        \includegraphics[height=1.75in]{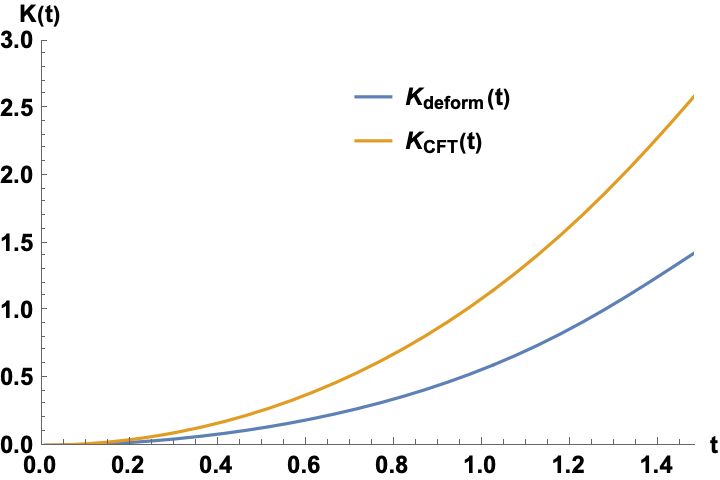}
        \caption{$\lambda=-0.08$}
    \end{subfigure}
    \hfill 
    \begin{subfigure}[h]{0.5\textwidth}
        \centering
        \includegraphics[height=1.75in]{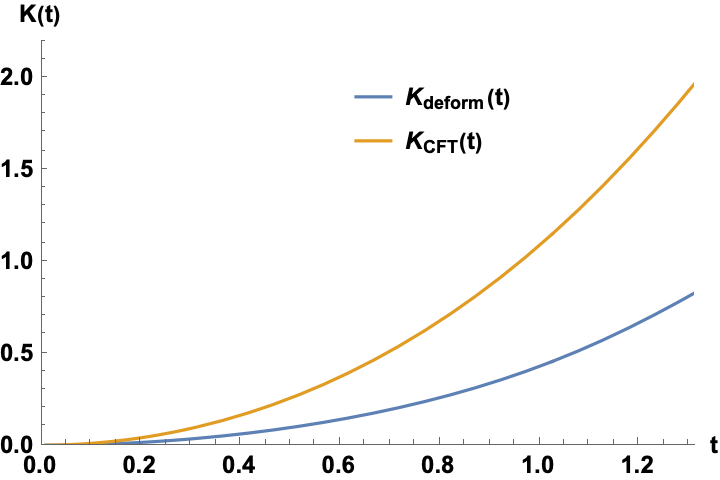}
        \caption{$\lambda=-0.10$}
    \end{subfigure}%
     \hfill 
    \begin{subfigure}[h]{0.5\textwidth}
        \centering
        \includegraphics[height=1.75in]{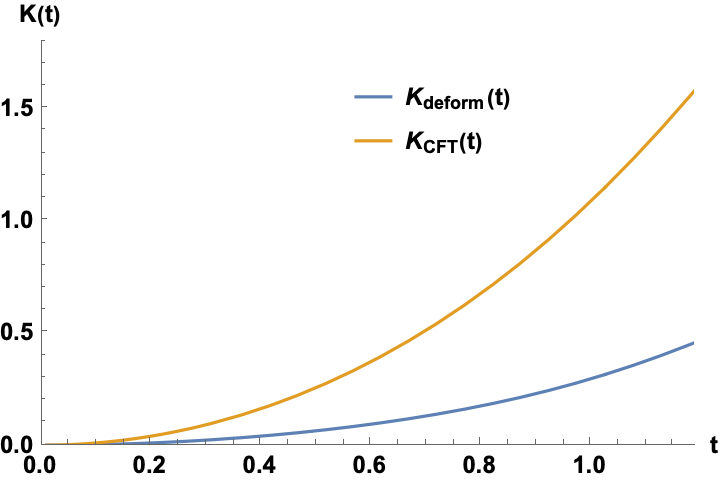}
        \caption{$\lambda=-0.12$}
    \end{subfigure}
     \hfill
   \caption{Krylov complexity of a two dimensional CFT contrasted with the the same post T$\bar{\rm T}$  deformation for various negative values of the deformation parameter $\lambda$. }
    \label{KCttbars}
\end{figure}
{In contrast to its counterpart discussed before, $b_n$'s do not go to zero for $\lambda<0$ in Fig. \ref{fig:TTbLanczosfig2}. One should still be cautious and observe that there is a  finite $n_{\text{max}}$ after which $O(\lambda)$ correction starts being of the same order as $O(\lambda^0)$. Furthermore, as discussed earlier because of the truncation of infinite sum corresponding to Krylov complexity, there is a time $t_c$ only up to  which the probabilities sum to unity (Fig. \ref{probttbars}) and there is a time $t_{\rm max}$ upto which the perturbative series for autocorrelation remains valid. Taking all these into account and sticking to the regime where $t_c>t_{\text{max}}$, Krylov complexity for this case is given in \cref{KCttbars}. Interestingly in this case $\lambda_K<{2\pi\over \beta}$ obeying the MSS bound. But as can be seen from \cref{fig:TTbLanczosfig2}, the growth of $b_n$ in this case is superlinear for $n>4$. Therefore, although with $\lambda<0$ we recover the MSS bound the price one had to pay is the superlinear growth of $b_n$. In fact, this observation is quite intriguing because of two interesting facts. The first one is that the growth of $b_n$ should be linearly bounded for local theories. On the other hand it has been observed that the asymptotic density of states for T$\bar{\rm T}$ deformed CFT with the "good" sign interpolates between Cardy behavior and Hagedorn behavior. The Hagedorn behavior indicates non-locality of T$\bar{\rm T}$ deformed CFTs. This exact observation fuelled the series of works in \cite{Giveon:2017nie, Giveon:2017myj, Giveon:2019fgr}, showing T$\bar{\rm T}$ deformed CFT as the boundary description of two dimensional vacua of little string theory. Therefore the superlinear behavior of the Lanczos coefficients of T$\bar{\rm T}$ deformed CFT for $\lambda<0$ is quite interesting but not so unexpected. 

 \begin{figure}[h!] 
    \centering
    \begin{subfigure}[h]{0.5\textwidth}
        \centering
        \includegraphics[height=1.9in]{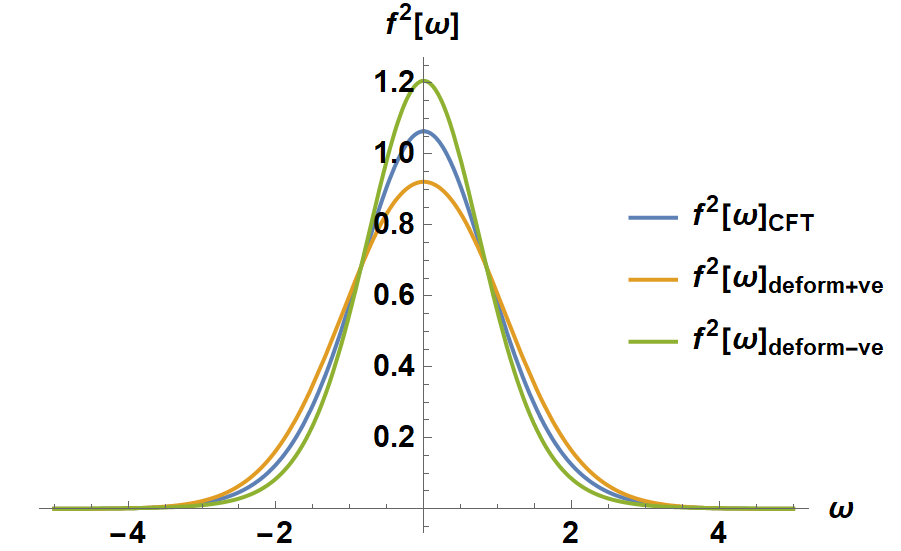}
        \caption{$\lambda=\pm 0.07$}
    \end{subfigure}%
    \hfill
    \begin{subfigure}[h]{0.5\textwidth}
        \centering
        \includegraphics[height=1.9in]{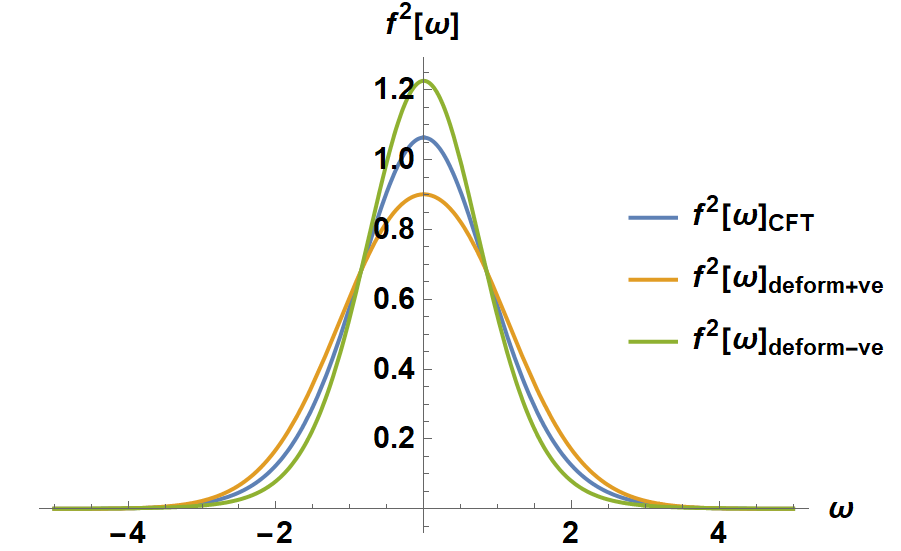}
        \caption{$\lambda=\pm 0.08$}
    \end{subfigure}
    \hfill 
    \begin{subfigure}[h]{0.5\textwidth}
        \centering
        \includegraphics[height=1.9in]{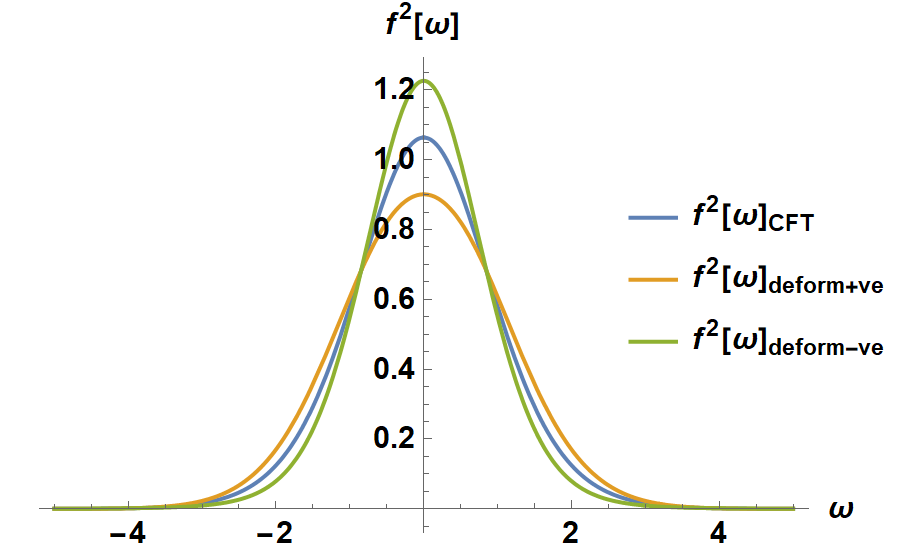}
        \caption{$\lambda=\pm 0.10$}
    \end{subfigure}%
     \hfill 
    \begin{subfigure}[h]{0.5\textwidth}
        \centering
        \includegraphics[height=1.9in]{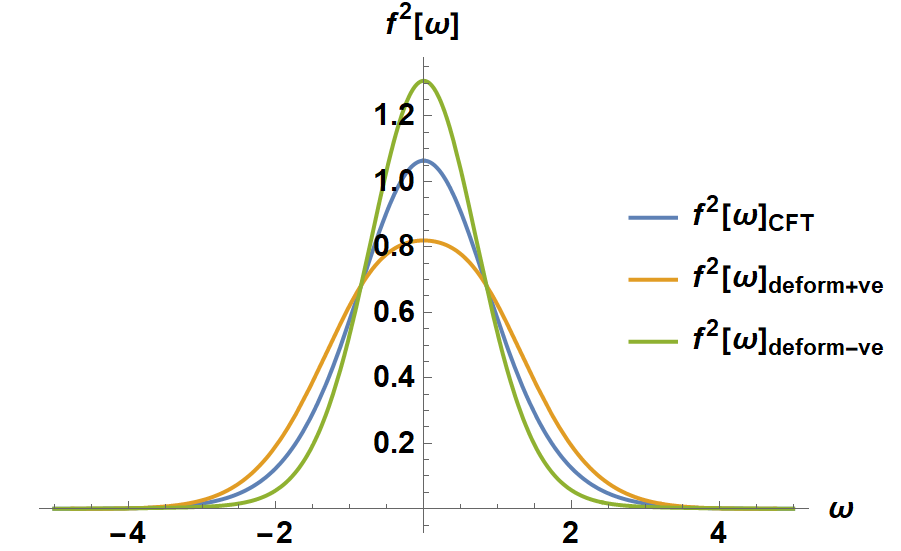}
        \caption{$\lambda=\pm 0.12$}
    \end{subfigure}
     \hfill
   \caption{Power spectrum of a two dimensional CFT contrasted with the the same post T$\bar{\rm T}$  deformation for both positive and negative values of the deformation parameter $\lambda$. }
    \label{fig:powerspectrum}
\end{figure}

\noindent Again we should emphasize that it can be a artifact of perturbative approach and to confirm this apparent superlinear growth one needs to compute higher order correction to the autocorrelation function\footnote{We thank Anatoly Dymarsky for drawing our attention to this point.}. In this regard we also studied the power spectrum by using \eqref{power}. Note that $C_0^{\lambda}(t)$ given in \eqref{autocorrelationTT} has both higher order poles and logarithmic branch points at $t=(2m+1)i\beta/2$ where $m=0,\pm 1\cdots$, compared to the undeformed case. Therefore one can expect corrections to the the asymptotic exponential tail of power spectrum discussed in \eqref{tail} originating from higher order poles and branch points. The analytical computations seems quite challenging and are beyond the scope of our current work. We have considered the power spectrum numerically for both the positive and negative values of the deformation parameter given in Fig. \ref{fig:powerspectrum}. One can observe that the power spectrum tail for positive deformation parameter is surpassing the undeformed CFT power spectrum tail which may support the apparent violation of universal operator growth hypothesis.

\subsection{J$\bar{\text{T}}$}
Following the same suit as before, we will now proceed with J$\bar{\rm T}$ deformation, picking up our computations for thermal two point function from \cref{JT:intresult}.
\subsubsection*{Autocorrelation function and Lanczos coefficients}
The autocorrelation function for this case can be obtained from \cref{JT:intresult} to give,
\begin{align}
C_0^\lambda(t)=\frac{1}{\cosh(\frac{\pi t }{\beta })^{2\Delta}}\bigg[1+ \Big(\frac{ \pi \lambda \Delta }{\beta}\Big)  	q\left(-1+\frac{\pi c}{6\Delta}\nu_2\right)\bigg]
\end{align}
Since the autocorrelation function is proportional to that of the undeformed CFT at the leading order in $\lambda$,
the  Lanczos coefficients $b_n$ corresponding to the above autocorrelation function receives no correction at this order and are given by
\begin{align}
    b(n)&=\frac{\pi  \sqrt{n (2 \Delta +n-1)}}{\beta }+{\cal O}(\lambda^2)\nonumber\\
\end{align}


\subsubsection*{Krylov complexity}
Since $b_n$s receive no correction at the leading order in $\lambda$ and the autocorrelation function is proportional to that of the undeformed CFT case, the wave functions are well known and the Krylov exponent is $\lambda_K=\frac{2\pi}{\beta}$, same as the undeformed CFT.

\subsection{J$\bar{\text{J}}$}
In this subsection, we determine the Lanczos coefficients and the Krylov complexity of an operator of a 2D CFT with J$\bar{\text{J}}$ deformation.

\subsubsection*{Autocorrelation function and Lanczos coefficients}
The autocorrelation function can be computed likewise,
\begin{align}
	C_0^\lambda(t)= \frac{1}{\Big(\cosh \frac{\pi t}{\beta}\Big)^{2\Delta}}\Big[1- {2\pi \lambda q^2  }\bigg(\nu_1+2\log\bigg(\cosh\frac{\pi t}{\beta}\bigg)\bigg)\bigg]+{\cal O}(\lambda^2)\nonumber\\
\end{align}
where $\nu_1=\left({2\over \epsilon}+\gamma+\ln\pi+2\ln 4\right)$. Which is computed through the result quoted in \cref{JJ:intresult}.
The corresponding  Lanczos Coefficients are as follows
\begin{align}
    b(n)&=\frac{\pi  \sqrt{n (2 \Delta +n-1)}}{\beta }+\lambda \bigg(\frac{ 2\pi ^2   q^2 \Delta }{\beta ^2}\bigg)\sqrt{\frac{n}{2 \Delta +n-1}}+{\cal O}(\lambda^2)\nonumber\\
\end{align}
As we can see for large $n$,
\begin{align}
    b_n\approx\alpha n+\delta
\end{align}
where the slope is $\alpha={\pi \over \beta}$ and hence unaffected by the deformation.

\subsubsection*{Krylov Complexity}
The necessary probability amplitudes are given by
\begin{align}
   |\phi(n,t)|^2=|\phi_{\rm undef}(n,t)|^2\bigg[1+ \lambda \frac{ 4\pi q^2 }{\beta } \left\{ \Delta  H_{n+2 \Delta -1}- \gamma  \Delta -\Delta  \psi ^{(0)}(2 \Delta )-2 \Delta  \log \left[\cosh \left(\frac{\pi  t}{\beta }\right)\right]\right\}\bigg]
\end{align}
where $|\phi_{\rm undef}(n,t)|$ correspond to the probability amplitudes for undeformed CFT and is given by
\begin{align}
  |\phi_{\rm undef}(n,t)|^2=  \frac{\Gamma (n+2 \Delta )}{\Gamma (2 \Delta ) \Gamma (n+1)}\tanh ^{2 n}\left(\frac{\pi  t}{\beta }\right) \text{sech}^{4 \Delta }\left(\frac{\pi  t}{\beta }\right)
\end{align}
The Krylov Complexity can be computed analytically in this case and is given by
\begin{align}
  K(t)= 2 \Delta\Big( 1 +\lambda \frac{2\pi q^2}{\beta }\Big)\sinh ^2\left(\frac{\pi  t}{\beta }\right)
\end{align}
which grows exponentially at long times with the undeformed Krylov exponent with $\lambda_K=\frac{2\pi}{\beta}$.
Hence unlike the T$\bar{\text{T}}$ case, here the Krylov complexity for a 2d CFT under J$\bar{\text{J}}$ deformation saturates the bound for Krylov exponent conjectured in \cite{Parker:2018yvk}.

\section{Discussion and Conclusion}\label{Sec:conclusion}
In this work, we have explored the behaviour of the Lanczos coefficients and the Krylov complexity for  T$\bar{\text{T}}$, J$\bar{\text{T}}$ and J$\bar{\text{J}}$ deformed two-dimensional CFTs perturbatively upto  first order of the deformation parameter. To this end, we began by first computing the thermal two point functions in these deformed theories by mapping the CFT on a Euclidean plane to a cylinder and then utilizing the dimensional regularization scheme to get a handle on the spacetime integrals. Subsequently, we determine the autocorrelation function which is related to the thermal two-point function through the Wightman inner product. Our final results show some intriguing features of these deformations. Before discussing the implication of these results, let us summarise the main observations 
\begin{itemize}
    \item \textbf{T$\bar{\text{T}}$ deformation:} In this case, we obtain the Lanczos coefficients, operator wave functions analytically. However we resort to numerical analysis for the Krylov complexity. Lanczos coefficients pick up $\mathcal{O}(\lambda)$ corrections and the Krylov exponent seemingly violates the universal bound proposed in  \cite{Avdoshkin:2022xuw} i.e $\lambda_K>{2\pi \over \beta}$. {For $\lambda<0$ we find that $\lambda_K<{2\pi \over \beta}$ but the Lanczos coefficients exhibit a superlinear growth violating the operator growth hypothesis \cite{Parker:2018yvk}.}
    \item \textbf{J$\bar{\text{T}}$ deformation:} Lanczos coefficients remain unaltered at  $\mathcal{O}(\lambda)$ and the Krylov exponent also remains same as that of the undeformed case saturating the conjectured bound.
    \item \textbf{J$\bar{\text{J}}$ deformation:} Although for small-$n$, the Lanczos coefficients receive correction at $\mathcal{O}(\lambda)$, their large-$n$ behaviour is same as that of the undeformed CFT. The Krylov complexity grows exponentially at long times while the Krylov exponent remains to be  ${2\pi \over \beta}$, which is exactly same as that of the undeformed CFT.
\end{itemize}
As we demonstrated, our observations are a result of perturbative expansion of autocorrelation functions as well as a careful truncation of the infinite  probability sum to a finite sum such that it still remains unity. However, to confirm our observation beyond the perturbative domain, one should either compute all higher order corrections to the Lanczos coefficients and resum or pursue a non-perturbative approach towards computing Krylov complexity for deformed CFTs. While J$\bar{\rm J}$ deformation can be analyzed non-perturbatively, it is quite challenging to get the exact non-perturbative thermal two-point function for T$\bar{\rm T}$ deformed theories \cite{Cardy:2019qao,Kruthoff:2020hsi}. Towards the first avenue, before computing the Lanczos coefficients one can naively expand the autocorrelation function to further higher orders of the deformation parameter $\lambda$ and subsequently observe their dependency on the central charge thus breaking universality. Furthermore, our computation of dimensionally regularized thermal two point function is evaluated up to $\mathcal{O}(\lambda)$ and therefore any $\mathcal{O}(\lambda^2)$ correction in the Lanczos coefficients cannot be presumably \emph{trusted} in the domain of this formalism. For a comprehensive understanding one has to modify the recursion relations involving the Lanczos coefficients accordingly. {We would like to point out that even though our observed violation of the extended MSS bound for T$\bar{\text{T}}$ deformed CFTs might apparently be seen as an artifact of our perturbation scheme, but our other observations with the same scheme of perturbation suggests otherwise. For the "good" sign of T$\bar{\text{T}}$ deformed CFT we observe a superlinear growth of $b_n$ reinforcing the fact of it being a non-local QFT, observed earlier in the literature \cite{Giveon:2017nie, Giveon:2017myj, Giveon:2019fgr}. Also we have seen no violation of the extended MSS bound for J$\bar{\text{T}}$ as well as J$\bar{\text{J}}$ deformed CFTs. Therefore the observed violation of MSS bound for T$\bar{\text{T}}$ deformed CFT with $\lambda>0$ deserves a more thorough investigation in the near future. }\vspace{1em}

\noindent It is worth mentioning that, in \cite{He:2019vzf} the authors proposed that T$\bar{\text{T}}$ deformed CFTs with large central charge, will not alter the  maximal chaos bound ($\lambda_L=\frac{2\pi}{\beta}$) identified by the exponential growth of OTOCs when treated perturbatively up to the first order in the deformation parameter. Furthermore, since the Lyapunov exponent derived from the OTOC remains unaffected, it was expected that the gravity dual of T$\bar{\text{T}}$ deformed CFTs will correspond to a theory which saturates the bound on chaos. One key assumption in \cite{He:2019vzf} is that the T$\bar{\text{T}}$ deformed theory retains its integrability up to the first order of deformation. Since our result apparently violates the extended MSS bound in certain region at this order, it may have interesting implications for its gravity dual \cite{Kraus:2018xrn, Hashemi:2019xeq, Katoch:2022hdf}. \vspace{1em}

\noindent Going beyond the first order correction in the deformation parameter for the Lanczos coefficients is quite challenging although it is much needed since we do not get any $\mathcal{O}(\lambda)$ corrections specially for J$\bar{\text{T}}$ as well as J$\bar{\text{J}}$ deformed theories. Therefore at present we are trying to develop a non-perturbative approach to compute Krylov complexity for field theories with deformations to confirm if the apparent violation of the conjectured bound for the Krylov exponent in the T$\bar{\text{T}}$ deformations case, is just an artifact of the limitations of our perturbative approach or it has some deeper implications. We hope to address this exciting issue in near future.

\section*{Acknowledgements}
 The work of AC is supported by the European Union Horizon 2020 research and innovation programme under the Marie Sk\l{}odowska Curie grant agreement number 101034383. The work of VM was supported  by the Brain Pool program funded by the Ministry of Science and ICT through the National Research Foundation of Korea (RS-2023-00261799). The work of AM is supported by POSTECH BK21 postdoctoral fellowship.  VM and AM acknowledge the support by the National Research Foundation of Korea (NRF) grant funded by the Korean government (MSIT) (No. 2022R1A2C1003182). AC acknowledges the hospitality of \"Orebro University during the final course of this work. We would like to thank Hugo A. Camargo, Viktor Jahnke, Mitsuhiro Nishida and Pantelis Panopoulos for useful discussions. We thank Anatoly Dymarsky for insightful comments on the draft.

\appendix

\section{Important integrals}\label{app:integrals}
In this section, we provide a comprehensive overview of the computational details employed to determine the integrals of thermal two point functions on cylinder for all three deformations.
\subsection{Integrals for T$\bar{\text{T}}$}\label{app:TTintegrals}

In the following we will explicitly evaluate $I_1$, 
\begin{align}
    I_1=|z_{12}|^{4h}\int dz d\bar{z} \, z \bar{z} \langle T(z)\bar{T}(\bar{z}) \mathcal{O}(z_1, \bar{z}_1) \mathcal{O}(z_2, \bar{z}_2)\rangle_0
\end{align}
We will state only the important steps of derivation for the rest. With some manipulations and disregarding contact terms, we have the relation \cite{He:2019vzf}
\begin{equation}
    \langle T(z)\bar{T}(\bar{z}) \mathcal{O}(z_1, \bar{z}_1) \mathcal{O}(z_1, \bar{z}_2)\rangle_{0}=h^2{|z_{12}|^4\over |z-z_1|^4|z-z_2|^4|z_{12}|^{4h}}.
\end{equation}
where we have assumed $h=\bar{h}$. In the following, we are going to evaluate the integrals utilizing the following Feynman parametrization,
\begin{eqnarray}
{1\over |z-z_1|^4|z-z_2|^4}=6\int_0^1 du{u(1-u)\over \left(|\tilde{z}|^2+u(1-u)|z_{12}|^2\right)^4}
\end{eqnarray}
where
\begin{align}
    z=\tilde{z}+u z_1+(1-u)z_2.
\end{align}
Therefore we have the relation
\begin{eqnarray}
    I_1&=&h^2|z_{12}|^4\int d^2 z  \frac{z \bar{z}}{|z-z_1|^4|z-z_2|^4}\notag\\
    &=&6 h^2 |z_{12}|^4\int_0^1 du\,u(1-u)\int {d^2\tilde{z} A(\tilde{z},\bar{\tilde{z}})\over \left(|\tilde{z}|^2+u(1-u)|z_{12}|^2\right)^4},
\end{eqnarray}
with $$A(\tilde{z},\bar{\tilde{z}})=|\tilde{z}|^2+\tilde{z}(u\bar{z}_{12}+\bar{z}_2)+\bar{\tilde{z}}(u z_{12}+z_2)+u|z_1|^2+(1-u)|z_2|^2-u(1-u)|z_{12}|^2.$$
One should note that the integration over the second and third term of $A(\tilde{z},\bar{\tilde{z}})$ will contribute to zero once we parametrize $\tilde{z}$ in polar coordinates as $\tilde{z}=\rho e^{i\theta}$. Therefore finally we have the relation
\begin{eqnarray}
    I_1=&&\overbrace{6 h^2 |z_{12}|^4\int_0^1du \, u(1-u)F_1^2(u)}^{T_1}+\overbrace{6 h^2|z_{12}|^4|z_1|^2\int_0^1du\, u^2(1-u)F_2^2(U)}^{T_2}\notag\\&&+\underbrace{6 h^2|z_{12}|^4|z_2|^2\int_0^1du\,u(1-u)^2 F_2^2(U)}_{T_3}
    -\underbrace{6 h^2|z_{12}|^6\int_0^1 du\, u^2(1-u)^2 F_2^2(U)}_{T_4},\notag\\
\end{eqnarray}
where $F_1^2(u), F_2^2(u)$ integrals are divergent.  Therefore we compute the integrals by dimensional regularization scheme according to
\begin{equation}
    F_1^d(u)=2 V_{s^{d-1}}\int_0^\infty {\rho^{d-1}\rho^2d\rho\over \left(\rho^2+u(1-u)|z_{12}|^2\right)^4};\quad F_2^d(u)=2 V_{s^{d-1}}\int_0^\infty {\rho^{d-1}d\rho\over \left(\rho^2+u(1-u)|z_{12}|^2\right)^4}.
\end{equation}
with $V_{s^{d-1}}={\pi^{d\over 2}\over \Gamma({d\over 2})}$.\vspace{1em}

\noindent Using the following relation 
\begin{equation}
    \int_0^\infty dt\,{t^{m-1}\over \left(t^2+a^2\right)^n}={1\over (a^2)^{n-{m\over 2}}}{\Gamma({m\over 2})\Gamma({n-{m\over 2})}\over 2 \Gamma({n})},
\end{equation}
we compute
\begin{eqnarray}
    F_1^d(u)={d \pi^{d/2}\over 12\left(|z_{12}|^2\right)^{3-d/2}}{\Gamma(3-{d\over 2})\over \left(u(1-u)\right)^{3-d/2}},\quad F_2^d(u)={\pi^{d/2}\over 6\left(|z_{12}|^2\right)^{4-d/2}}{\Gamma(4-{d\over 2})\over \left(u(1-u)\right)^{4-d/2}}.
\end{eqnarray}
Finally we can carry out the integration over $u$ in $T_1$ using the following integral,
\begin{equation}
    \int_0^1{du\over \left(u(1-u)\right)^{2-d/2}}={2^{3-d}\sqrt{\pi}\Gamma\left({d\over 2}-1\right)\over \Gamma({d\over 2}-{1\over 2})},
\end{equation}
yielding
\begin{equation}\label{T1}
    T_1={d\over 2}h^2\pi^{d/2}\left(|z_{12}|^2\right)^{{d\over 2}-1}\Gamma\left(3-{d\over 2}\right){2^{3-d}\sqrt{\pi}\Gamma\left({d\over 2}-1\right)\over \Gamma({d\over 2}-{1\over 2})}
\end{equation}
This converges for $\text{Re}(d)>2$ but can be analytically continued to any complex value of $d$. Therefore we can now take the limiting case $d=2+\epsilon$ in \cref{T1}, with $\epsilon$ being the regularization parameter,
\begin{equation}\label{T1n}
    T_1=\mu^{\epsilon\over 2}\left(1+{\epsilon\over 2}\right)\,h^2\,\pi \pi^{\epsilon/2}\left(|z_{12}|^2\right)^{\epsilon/2}\left(1-{\epsilon\over 2}\right){\Gamma(1-{\epsilon\over 2})\Gamma({\epsilon\over 2})\over \Gamma({\epsilon\over 2})\Gamma({1\over 2}+{\epsilon\over 2})}2^{1-\epsilon}\sqrt{\pi}\,\Gamma({\epsilon\over 2}). 
\end{equation}
where $\mu$ is introduced as an arbitrary mass scale in order to keep the correct units for the integral under dimensional regularisation. Furthermore one can use the properties of Gamma functions 
$$\begin{matrix}\Gamma(1-z)\Gamma(z)=\displaystyle{\pi\over \sin \pi z}\quad \text{Euler's reflection formula}\\ \\\Gamma(z)\Gamma(z+{1\over 2})=2^{1-2 z}\sqrt{\pi}\,\Gamma(2 z)\quad \text{Legendre duplicate formula}\end{matrix}$$
$$\lim_{x\rightarrow 0}\Gamma(x)={1\over x}-\gamma+{6 \gamma^2+\pi^2\over 12}x+\mathcal{O}(x^2)$$
in \cref{T1n} to reduce
\begin{eqnarray}
    T_1=2 \pi h^2\left({2\over \epsilon}+\gamma+\ln \pi+\ln|z_{12}|^2+\ln\mu\right)+\mathcal{O}(\epsilon).
\end{eqnarray}
To compute $T_2$ and $T_3$ we require the following two integrals which give same results,
\begin{equation}
    \int_0^1 {u^2(1-u)\over \left(u(1-u)\right)^{4-{d\over 2}}}du=\int_0^1 {u(1-u)^2\over \left(u(1-u)\right)^{4-{d\over 2}}}du={2^{4-d}\sqrt{\pi}\,\Gamma({d\over 2}-2)\over \Gamma({d\over 2}-{3\over 2})}
\end{equation}
Using the same approach as before and analytically continue to any complex value for $d$ and using $d=2+\epsilon$ as before we have
\begin{eqnarray}
    T_2&=&|z_1|^2 h^2{2\pi\over |z_{12}|^2}\left({4\over \epsilon}+2\gamma-5+2\ln \pi+2\ln|z_{12}|^2+\ln\mu^2\right)+\mathcal{O}(\epsilon),\\
    T_3&=&|z_2|^2 h^2{2\pi\over |z_{12}|^2}\left({4\over \epsilon}+2\gamma-5+2\ln \pi+2\ln|z_{12}|^2+\ln\mu^2\right)+\mathcal{O}(\epsilon).
\end{eqnarray}
In a similar manner, the following integral
\begin{align}
    \int_0^1 \frac{u^2 (1-u)^2}{(u (1-u))^{4-\frac{d}{2}}}=\frac{\Gamma \left(\frac{d}{2}-1\right)^2}{\Gamma (d-2)}
\end{align}
yields
\begin{equation}
    T_4=2 \pi \,h^2\left({4\over \epsilon}+2\gamma-3+2\ln\pi+2\ln|z_{12}|^2+2\ln\mu\right)+\mathcal{O}(\epsilon).
\end{equation}
Hence we finally have the final result for $I_1$ by adding all contributions from $T_i$'s with (i=1 to 4),
\begin{eqnarray}
    I_1
    &=& 2\pi h^2{|z_1|^2 +|z_2|^2\over |z_{12}|^2}\left({4\over \epsilon}+2\gamma-5+2\ln \pi+2\ln|z_{12}|^2+\ln\mu^2\right)\notag\\
    &&-2\pi h^2\left({2\over \epsilon}+\gamma-3+\ln\pi+\ln|z_{12}|^2+\ln\mu\right)+\mathcal{O}(\epsilon).
\end{eqnarray}
Next we will evaluate 
\begin{align}
    I_2&=-\frac{c |z_{12}|^{4h}}{24 }\int  dz d\bar{z} \, \frac{z}{\bar{z}}\langle T(z) \mathcal{O}(z_1, \bar{z}_1) \mathcal{O}(z_2, \bar{z}_2)\rangle_0\notag\\&={-ch}\int  dz d\bar{z} \, \frac{z}{\bar{z}}\left[\frac{1}{(z-z_1)^2}+\frac{1}{(z-z_2)^2}-{2\over {(z-z_1)z_{12}}}+{2\over {(z-z_2)z_{12}}}\right]
\end{align}
Likewise
\begin{align}
I_3&
={-ch}\int  dz d\bar{z} \, \frac{\bar{z}}{z}\left[\frac{1}{(\bar{z}-\bar{z}_1)^2}+\frac{1}{(\bar{z}-\bar{z}_2)^2}-{2\over {(\bar{z}-\bar{z}_1)\bar{z}_{12}}}+{2\over {(\bar{z}-\bar{z}_2)\bar{z}_{12}}}\right]
\end{align}
Before tackling these integrals, let us focus on the following,
\begin{eqnarray}\label{eq:I0}
    \mathcal{I}^0(z_i)=\int dzd\bar{z}\,{z\over \bar{z}(z-z_i)}&=&\int dzd\bar{z}\,{z^2(\bar{z}-\bar{z}_i)\over |z|^2|z-z_i|^2}\notag\\
    &=&\int_0^1du\int d^2\hat{z}{B(\hat{z})\over \left(|\hat{z}|^2+u(1-u)|z_i|^2\right)^2},
\end{eqnarray}
where $\hat{z}=z-(1-u)z_i$ and 
\begin{eqnarray}
B(\hat{z})=z^2(\bar{z}-\bar{z}_i)=&2z_i(1-u)|\hat{z}|^2-u(1-u)^2|z_i|^2z_i+\bar{\hat{z}}z_i^2(1-u)^2+|\hat{z}|^2\hat{z}\notag\\
&-\hat{z}^2u\bar{z}_i-2|z_i|^2\hat{z}u(1-u).
\end{eqnarray}
One should note that only the first two terms will contribute to the integral with the polar coordinate parametrization, therefore we have
\begin{align}
\mathcal{I}^0(z_i)=&4 z_i V_{s^{2-1}}\int_0^1 du\,(1-u)\int_0^\infty {\rho^{2-1}\rho^2d\rho\over\left(\rho^2+u(1-u)|z_i|^2\right)^2}\notag\\
&-2 |z_i|^2z_i V_{s^{2-1}}\int_0^1 du\,u(1-u)^2\int_0^\infty {\rho^{2-1}d\rho\over\left(\rho^2+u(1-u)|z_i|^2\right)^2} 
\end{align}
This can be regularized through the same procedure as before to have
\begin{eqnarray}
    \mathcal{I}^0(z_i)={\pi z_i}\left({1\over2}-{2\over \epsilon}-\gamma-\ln \pi-\ln|z_i|^2-\ln\mu\right)+\mathcal{O}(\epsilon)
\end{eqnarray}
One can easily find that $I_2$ and $I_3$ can be written as the derivatives of $\mathcal{I}^0$
\begin{eqnarray}
    I_2&=-{c\,h\over 24}\left[\partial_{z_1}\mathcal{I}^0(z_1)+\partial_{z_2}\mathcal{I}^0(z_2)-{2\over z_{12}}\mathcal{I}^0(z_1)+{2\over z_{12}}\mathcal{I}^0(z_2)\right]\notag\\I_3&=-{c\,h\over 24}\left[\partial_{\bar{z}_1}\mathcal{I}^0(\bar{z}_1)+\partial_{\bar{z}_2}\mathcal{I}^0(\bar{z}_2)-{2\over \bar{z}_{12}}\mathcal{I}^0(\bar{z}_1)+{2\over \bar{z}_{12}}\mathcal{I}^0(\bar{z}_2)\right].
\end{eqnarray}
Hence the final results are,
\begin{eqnarray}
I_2={c\,h\over 24}\left[2 \pi+\pi \ln|z_1 z_2|^2-{2\pi\over z_{12}}\left(z_1\ln|z_1|^2-z_2\ln|z_2|^2\right)\right]=\bar{I}_3
\end{eqnarray}

\subsection{Integrals for J$\bar{\text{T}}$}\label{JTint}\label{app:jtbar}
Now we will consider the integrals required for the computation of thermal two point function of J$\bar{\text{T}}$ deformed CFT.
\begin{align}
\mathbb{I}_1(z_1,\bar{z}_1;z_2,\bar{z}_2)&=|z_{12}|^{4h}\int dz d\bar{z} \, {\bar{z}} \langle J(z)\bar{T}(\bar{z}) \mathcal{O}(z_1, \bar{z}_1) \mathcal{O}(z_2, \bar{z}_2)\rangle_{0}\\
\mathbb{I}_2(z_1,\bar{z}_1;z_2,\bar{z}_2)&=-\frac{c |z_{12}|^{4h}}{24 }\int  dz d\bar{z} \, {1\over \bar{z}}\langle J(z) \mathcal{O}(z_1, \bar{z}_1) \mathcal{O}(z_2, \bar{z}_2)\rangle_{0}
\end{align}
We split $\mathbb{I}_1$ in two parts,
\begin{align}\label{eq:I1form}
    \mathbb{I}_1=\mathbb{K}_1+\mathbb{K}_2
\end{align}
where
\begin{align}
    \label{eq:j1j2-1}
    \mathbb{K}_1&=\lambda q_1\bar{h}\bar{z}^2_{12} \int dz d\bar{z} \, \frac{\bar{z}}{(z-z_1)(\bar{z}-\bar{z}_1)^2(\bar{z}-\bar{z}_2)^2}\\
    \label{eq:j1j2-2}
    \mathbb{K}_2&=\lambda q_2\bar{h}\bar{z}^2_{12} \int dz d\bar{z} \, \frac{\bar{z}}{(z-z_2)(\bar{z}-\bar{z}_1)^2(\bar{z}-\bar{z}_2)^2} 
\end{align}
Plugging the OPE of $\text{J}$ and ${\cal O}$, $\mathbb{I}_2$ boils down to
\begin{align}\label{eq:i2}
   \mathbb{I}_2=-\frac{c}{24 }\int  dz d\bar{z} \, \frac{1}{\bar{z}}\left(\frac{q_1}{z-z_1}+\frac{q_2}{z-z_2}\right)
\end{align}
To solve \eqref{eq:j1j2-1}, \eqref{eq:j1j2-2} and \eqref{eq:i2}, we would first focus on the following four integrals
\begin{eqnarray}
    \mathcal{I}^1(z_1,\bar{z}_1)&=&\int dzd\bar{z}\,{1\over \bar{z}(z-z_1)}\\
    \mathcal{I}^2 (z_1,\bar{z}_1,z_2,\bar{z}_2)&=&\int dzd\bar{z}\,{\bar{z}\over (z-z_1)(\bar{z}-\bar{z}_2)}\\
    \mathcal{I}^3(z_1,\bar{z}_1)&=&\int dz d\bar{z}{\bar{z}\over |z-z_1|^2}\\
    \mathcal{I}^4(z_1,\bar{z}_1,z_2,\bar{z}_2)&=&\int dz d\bar{z}\,{\bar{z}\over (z-z_1)(\bar{z}-\bar{z}_1)(\bar{z}-\bar{z}_2)}
\end{eqnarray}
\addtocontents{toc}{\protect\setcounter{tocdepth}{1}}
\subsubsection{$\mathcal{I}^1(z_1,\bar{z}_1)$}
We will start by rewriting the integral as
\begin{eqnarray}
    \int dz d\bar{z}{1\over \bar{z}(z-z_1)}&=&\int dz d\bar{z}{z(\bar{z}-\bar{z}_1)\over |z|^2|z-z_1|^2}\notag\\
    &=&\int_0^1 du\int d\hat{z}{C(\hat{z})\over\left(|\hat{z}|^2+u(1-u)|z_1|^2)\right)^2}
\end{eqnarray}
where $\hat{z}=z-(1-u)z_1$ and 
$C(\hat{z})=|\hat{z}|^2-u(1-u)|z_1|^2-u z_1\hat{z}+(1-u)z_1\bar{\hat{z}} $\vspace{1em}\\
Only the first two terms in $C(\hat{z})$ will contribute to the integral with the polar coordinate parametrization. Therefore this integral can be broken up in the following two terms
\begin{eqnarray}
    \mathcal{I}^1(z_1,\bar{z}_1)&=&\overbrace{2 V_{S^2-1}\int_0^1du\,\int_0^\infty{\rho^{2-1}\rho^2 d\rho\over \left(\rho^2+u(1-u)|z_{1}|^2\right)^2}}^{U_1}\notag\\
    && -\overbrace{2 V_{S^2-1}|z_{1}|^2\int_0^1du\,u(1-u)\int_0^\infty{\rho^{2-1}d\rho\over \left(\rho^2+u(1-u)|z_{1}|^2\right)^2}}^{U_2}
\end{eqnarray}
Using the dimensional regularization techniques we have the simple relation upto $\mathcal{O}(\epsilon)$ that
$$U_1=-\pi\left({2\over \epsilon}-1+\gamma+\ln\pi+\ln|z_1|^2\right),\quad U_2=\pi.$$
which finally results in,
\begin{equation}
    \mathcal{I}^1(z_1,\bar{z}_1)=-\pi\left({2\over \epsilon}+\gamma+\ln\pi+\ln|z_1|^2+\ln \mu\right)+\mathcal{O}(\epsilon)
\end{equation}
\subsubsection{$\mathcal{I}^2(z_1,\bar{z}_1,z_2,\bar{z}_2)$}
In this case, we can first introduce the Feynman parametrization as the following
\begin{eqnarray}
    \int dzd\bar{z}\,{\bar{z}\over (z-z_1)(\bar{z}-\bar{z_2})}&=&\int dzd\bar{z}\,{\bar{z}(\bar{z}-\bar{z}_1)(z-z_2)\over |z-z_1|^2|z-z_2|^2}\nonumber\\
    &=&\int_0^1 du\int d^2\tilde{z}{D(\tilde{z})\over \left(|\tilde{z}|^2+u(1-u)|z_{12}|^2\right)^2}
\end{eqnarray}
where 
\begin{eqnarray}
    D(\tilde{z})&=&u^2(u-1)|z_{12}|^2\bar{z}_{12}+u(u-1)|z_{12}|^2\bar{z}_{2}+(2u-1)\bar{z}_{12}|\tilde{z}|^2+\bar{z}_2|\tilde{z}|^2+\cdots
\end{eqnarray}
with the definition $z=\tilde{z}+uz_1+(1-u)z_2$. One should note that all the rest of the terms in $\cdots$ in the expression of $\bar{z}(\bar{z}-\bar{z}_1)(z-z_2)$ will integrate to zero. Therefore one can break the integral for convenience as per the following,
\begin{eqnarray}
\mathcal{I}^2(z_1,\bar{z}_1,z_2,\bar{z}_2)&=&\overbrace{2 V_{S^2-1}|z_{12}|^2\bar{z}_{12}\int_0^1du\,u^2(u-1)\int_0^\infty{\rho^{2-1}d\rho\over \left(\rho^2+u(1-u)|z_{12}|^2\right)^2}}^{\mathbb{T}_1}\notag\\
&& +\overbrace{2 V_{S^2-1}|z_{12}|^2\bar{z}_{2}\int_0^1du\,u(u-1)\int_0^\infty{\rho^{2-1}d\rho\over \left(\rho^2+u(1-u)|z_{12}|^2\right)^2}}^{\mathbb{T}_2}\notag\\
 && +\overbrace{2 V_{S^2-1}\bar{z}_{12}\int_0^1du\,(2u-1)\int_0^\infty{\rho^{2-1}\rho^2d\rho\over \left(\rho^2+u(1-u)|z_{12}|^2\right)^2}}^{\mathbb{T}_3}\notag\\
    && +\overbrace{2 V_{S^2-1}\bar{z}_{2}\int_0^1du\int_0^\infty{\rho^{2-1}\rho^2d\rho\over \left(\rho^2+u(1-u)|z_{12}|^2\right)^2}}^{\mathbb{T}_4}.
\end{eqnarray}
Following the same steps as before for dimensional regularization we have the following result upto $\mathcal{O}(\epsilon)$
\begin{equation}
    \begin{split}
       \mathbb{T}_1&=-{\pi\over 2}\bar{z}_{12},\quad  \mathbb{T}_2=-\pi \bar{z}_2 ,\quad \mathbb{T}_3=0\\
       \mathbb{T}_4&=-\pi \bar{z}_2\left({2\over \epsilon}-1+\gamma+\ln \pi+\ln|z_{12}|^2\right).
    \end{split}
\end{equation}
Collecting all these together we have
\begin{equation}
    \mathcal{I}^2(z_1,\bar{z}_1,z_2,\bar{z}_2)=-{\pi\over 2}\bar{z}_{12}-\pi \bar{z}_2\left({2\over \epsilon}+\gamma+\ln \pi+\ln|z_{12}|^2+\ln \mu\right)+\mathcal{O}(\epsilon).
\end{equation}
\subsubsection{$\mathcal{I}^3(z_1,\bar{z}_1)$}
To get the results of this integral in our preferred format we will rewrite the integral as
\begin{eqnarray}
    \int dz d\bar{z}{\bar{z}\over |z-z_1|^2}&=&\int dz d\bar{z} {\bar{z}|z|^2\over |z|^2|z-z_1|^2}\notag\\
    &=&\int_0^1 du\int d^2\hat{z} {E(\hat{z})\over \left(|\hat{z}|^2+u(1-u)|z_1|^2\right)^2}
\end{eqnarray}
where $\hat{z}=z-(1-u)z_1$ and $E(\hat{z})=(1-u)^3|z_1|^2\bar{z}_1+2(1-u)\bar{z}_1|\hat{z}|^2$ with all other terms finally integrate to zero. Hence we have the relation
\begin{eqnarray}
    \int dz d\bar{z}{\bar{z}\over |z-z_1|^2}&=&\overbrace{2 V_{S^{2-1}}|z_1|^2\bar{z}_1\int_0^1 du\,(1-u)^3\int_0^\infty d\rho{\rho^{2-1}d\rho\over \left(\rho^2+u(1-u)|z_{1}|^2\right)^2}}^{\mathbb{T}_5}\notag\\
    &&+\overbrace{4 V_{S^{2-1}}\bar{z}_1\int_0^1 du\,(1-u)\int_0^\infty d\rho{\rho^{2-1}\rho^2d\rho\over \left(\rho^2+u(1-u)|z_{1}|^2\right)^2}}^{\mathbb{T}_6}
\end{eqnarray}
Again utilizing the dimensional regularization process we obtain, 
$$\mathbb{T}_5=\pi \bar{z}_1\left({2\over \epsilon}-{3\over 2}+\gamma+\ln \pi+\ln|z_1|^2\right),\quad \mathbb{T}_6=-{\pi}\bar{z}_1\left({2\over \epsilon}-1+\gamma+\ln\pi+\ln|z_1|^2\right).$$
Therefore we have the final result
{\begin{equation}
    \mathcal{I}^3(z_1,\bar{z}_1)=-{1\over 2}{\pi}\bar{z}_1+\mathcal{O}(\epsilon)
\end{equation}}
\subsubsection{$\mathcal{I}^4(z_1,\bar{z}_1,z_2,\bar{z}_2)$}
We can rewrite $\mathcal{I}^4$ according to,
\begin{eqnarray}
    \mathcal{I}^4(z_1,\bar{z}_1,z_2,\bar{z}_2)&=&{1\over \bar{z}_{12}}\int dz d\bar{z}\left({\bar{z}\over |z-z_1|^2}-{\bar{z}\over{(z-z_1)(\bar{z}-\bar{z}_2)}}\right)\notag\\
    &=&{1\over \bar{z}_{12}}\left[\mathcal{I}^3(z_1,\bar{z}_1)-\mathcal{I}^2(z_1,\bar{z}_1,z_2,\bar{z}_2)\right]
\end{eqnarray}
by plugging
\begin{eqnarray}
    {1\over (\bar{z}-\bar{z}_1)(\bar{z}-\bar{z}_2)}&=&{1\over z_{12}}\left({1\over \bar{z}-\bar{z}_1}-{1\over \bar{z}-\bar{z}_2}\right).
\end{eqnarray}
Therefore, we finally have
\begin{eqnarray}
    &&\mathcal{I}^4(z_1,\bar{z}_1,z_2,\bar{z}_2)=
    {\pi\over \bar{z}_{12}}\left[-{\bar{z}_1\over 2}+{\bar{z}_{12}\over 2}+{\bar{z}_2}\left({2\over \epsilon}+\gamma+\ln\pi+\ln|z_{12}|^2+\ln \mu\right)\right]+\mathcal{O}(\epsilon)\notag\\
\end{eqnarray}
\addtocontents{toc}{\protect\setcounter{tocdepth}{2}}
\noindent With all these calculations we now have the simple relation
\begin{eqnarray}\label{eq:J1res}
\mathbb{K}_1&=&\lambda q_1\bar{h}\bar{z}_{12}^2\int dz d\bar{z}{\bar{z}\over (z-z_1)(\bar{z}-\bar{z}_1)^2(\bar{z}-\bar{z}_2)^2}=\lambda q_1\bar{h}\bar{z}_{12}^2\,\partial_{\bar{z}_1}\partial_{\bar{z_2}}\mathcal{I}^4(z_1,\bar{z}_1,z_2,\bar{z}_2)\notag\\
&=&-{\pi\over 2\bar{z}_{12}}(\bar{z}_1+\bar{z}_2)\lambda q_1\bar{h}\left({4\over \epsilon}-5+2\gamma+2\ln\pi+2\ln|z_{12}|^2+\ln \mu^2\right)-{\pi\over \bar{z}_{12}}\lambda q_1 \bar{h}\bar{z}_1\notag\\
\end{eqnarray}
A similar analysis will yield
\begin{eqnarray}\label{eq:J2res}
\mathbb{K}_2&=&{\pi\over 2\bar{z}_{12}}(\bar{z}_1+\bar{z}_2)\lambda q_2\bar{h}\left({4\over \epsilon}-5+2\gamma+2\ln\pi+2\ln|z_{12}|^2+\ln \mu^2\right)+{\pi\over \bar{z}_{12}}\lambda q_2 \bar{h}\bar{z}_2\notag\\
\end{eqnarray}
A simple addition of \eqref{eq:J1res} and \eqref{eq:J2res} will generate the integral $I_1$ in \eqref{eq:I1form} upto $\mathcal{O}(\epsilon)$. 
\begin{align}
\mathbb{I}_1&={\pi\lambda \bar{h} \over 2\bar{z}_{12}}(\bar{z}_1+\bar{z}_2) (q_2-q_1)\left({4\over \epsilon}-5+2\gamma+2\ln\pi+2\ln|z_{12}|^2+\ln \mu^2\right)\notag\\&\qquad\qquad\qquad+{\pi\over \bar{z}_{12}}\lambda\, \bar{h}\, (q_2 \bar{z}_2-q_1 \bar{z}_1)
\end{align}
We can evaluate $\mathbb{I}_2$ as follows,
\begin{eqnarray}
\mathbb{I}_2&=&-\frac{c}{24 }\int  dz d\bar{z} \, \frac{1}{\bar{z}}\left(\frac{q_1}{z-z_1}+\frac{q_2}{z-z_2}\right)\notag\\
&=&-{c\over 24}\left(q_1\mathcal{I}^1(z_1,\bar{z}_1)+q_2\mathcal{I}^1(z_2,\bar{z}_2)\right)\notag\\&=&{\pi c\over 24}\left[\left(q_1+q_2\right)\left({2\over \epsilon}+\gamma+\ln\pi+\ln \mu\right)+q_1\ln|z_1|^2+q_2\ln|z_2|^2\right]
\end{eqnarray}
\subsection{Integrals for J$\bar{\text{J}}$}\label{app:jjbar}
In this case, we need to evaluate the following two integrals
\begin{eqnarray}
    \mathcal{I}^5(z_1,\bar{z}_1)&=&\int {dz d\bar{z}\over |z-z_1|^2}\\
    \mathcal{I}^6(z_1,\bar{z}_1,z_2,\bar{z}_2)&=&\int {dzd\bar{z}\over (z-z_1)(\bar{z}-\bar{z}_2)}
\end{eqnarray}

\addtocontents{toc}{\protect\setcounter{tocdepth}{1}}
\subsubsection{$\mathcal{I}^5(z_1,\bar{z}_1)$}
We rewrite $\mathcal{I}^5$ in the following way,
\begin{eqnarray}
    \int {dz d\bar{z}\over |z-z_1|^2}&=&\int dz d\bar{z}{|z|^2\over |z|^2|z-z_1|^2}\notag\\
    &=&\int_0^1 du\int d^2\hat{z}{E(\hat{z})\over \left(|\hat{z}|^2+u(1-u)|z_1|^2\right)^2}
\end{eqnarray}
with $\hat{z}=z-(1-u)z_1$, and ~~$E(\hat{z})=z\bar{z}=|\hat{z}|^2+(1-u)^2|z_1|^2+(1-u)(\bar{z}_1\hat{z}+z_1\bar{\hat{z}}).$\\
Only the first two terms in $E(\hat{z})$ will contribute in the integral, therefore we have
\begin{eqnarray}
    \int {dz d\bar{z}\over |z-z_1|^2}&=& \overbrace{2V_{S^{2-1}}\int_0^1 du\int_0^\infty {\rho^{2-1}\rho^2d\rho\over \left(\rho^2+u(1-u)|z_1|^2\right)^2}}^{{\cal{T}}_1}\notag\\
    &&+\overbrace{2 V_{S^{2-1}}|z_1|^2\int_0^1 du (1-u)^2\int_0^\infty {\rho^{2-1}d\rho\over \left(\rho^2+u(1-u)|z_1|^2\right)^2}}^{\mathcal{T}_2}
\end{eqnarray}
These integrals leads to,
\begin{equation}
    \begin{split}
       \mathcal{T}_1=-\pi\left({2\over \epsilon}-1+\gamma+\ln\pi+\ln|z_1|^2\right)+\mathcal{O}(\epsilon)\\
       \mathcal{T}_2=\pi\left({2\over \epsilon}-1+\gamma+\ln\pi+\ln|z_1|^2\right)+\mathcal{O}(\epsilon),
    \end{split}
\end{equation}
finally contributing to
\begin{equation}
    \mathcal{I}^5(z_1,\bar{z}_1)= \mathcal{O}(\epsilon)
\end{equation}
\subsubsection{$\mathcal{I}^6(z_1,\bar{z}_1,z_2,\bar{z}_2$)}
Similar to the earlier integrals, in this case we will rewrite the integral likewise,
\begin{eqnarray}
    \int {dz d\bar{z}\over (z-z_1)(\bar{z}-\bar{z}_2)}&=&\int dzd\bar{z}{(\bar{z}-\bar{z}_1)(z-z_2)\over |z-z_1|^2|z-z_2|^2}\notag\\
    &=&\int_0^1du\int d^2\tilde{z}{F(\tilde{z})\over \left(|\tilde{z}|^2+u(1-u)|z_{12}|^2\right)^2}
\end{eqnarray}
where as usual $z=\tilde{z}+uz_1+(1-u)z_2$ and $$F(\tilde{z})=|\tilde{z}|^2-u(1-u)|z_{12}|^2+\cdots$$
where "$\cdots$" denotes other terms which will not contribute to this integral. Hence
\begin{eqnarray}
    \int {dz d\bar{z}\over (z-z_1)(\bar{z}-\bar{z}_2)}&=&\overbrace{2 V_{S^{2-1}}\int_0^1 du\int_0^\infty {\rho^{2-1}\rho^2d\rho\over\left(\rho^2+u(1-u)|z_{12}|^2\right)^2}}^{\mathscr{T}_3}\notag\\
    &&-\overbrace{2 V_{S^{2-1}}|z_{12}|^2\int_0^\infty du\, u(1-u)\int_0^\infty {\rho^{2-1}d\rho\over\left(\rho^2+u(1-u)|z_{12}|^2\right)^2}}^{\mathscr{T}_4}\notag\\
\end{eqnarray}
Carrying out the dimensional regularization, we obtain
$$\mathscr{T}_3=-\pi\left({2\over \epsilon}-1+\gamma+\ln\pi+\ln|z_{12}|^2\right),\quad\mathscr{T}_4=\pi,$$
which furthermore gives,
\begin{equation}
    \mathcal{I}^6(z_1,\bar{z}_1,z_2,\bar{z}_2)=-\pi\left({2\over \epsilon}+\gamma+\ln\pi+\ln|z_{12}|^2+\ln \mu\right)+\mathcal{O}(\epsilon)
\end{equation}

\noindent Collecting all these calculations together we finally have 
\begin{eqnarray}
    &&{\cal Y}_1+{\cal Y}_2+{\cal Y}_3+{\cal Y}_4\notag\\
    &=& |q_1|^2\,\mathcal{I}^5(z_1,\bar{z}_1)+ |q_2|^2\,   \mathcal{I}^5 (z_2,\bar{z}_2)+q_1\bar{q}_2\,\mathcal{I}^6(z_1,\bar{z}_1,z_2,\bar{z}_2)+\bar{q}_1q_2\,\mathcal{I}^6(z_2,\bar{z}_2,z_1,\bar{z}_1)\notag\\
    &=&-\pi\left(q_1\bar{q}_2+\bar{q}_1q_2\right)\left({2\over \epsilon}+\gamma+\ln\pi+\ln|z_{12}|^2+\ln \mu\right)+\mathcal{O}(\epsilon).
\end{eqnarray}

\section{$|\phi_n(t)|^2$ for T$\bar{\text{T}}$}\label{OPAT}

 The probabilities corresponding to the operator wave functions are as follows
\begin{align}
	|\phi_n(t)|^2=P_0(n,t)-\tilde{\lambda}\{P_1(n,t)+\nu_1P_2(n,t)\}+{\cal O}(\lambda^2)
\end{align}
The functions $P_0,P_1,P_2$ in the above expression are given as follows
\begin{align*}
	P_0(n,t)&=\frac{1}{6} (n+1) (n+2) (n+3) \text{sech}^8\left(\frac{t}{2}\right) \tanh ^{2 n}\left(\frac{t}{2}\right)\\
	P_1(n,t)&=	\pi f_0^2(n,t)\bigg[g_1(n,t)+	g_2(n,t)+	g_3(n,t)+g_4(n,t)+	g_5(n,t)+	g_6(n,t)+g_7(n,t)\nonumber\\& \qquad +	g_8(n,t)+	g_9(n,t)+g_{10}(n,t)+	g_{11}(n,t)+	g_{12}(n,t)\bigg]\\
	P_2(n,t)&=\pi f_0^2(n,t)\bigg[h_1(n,t)+	h_2(n,t)+	h_3(n,t)+h_4(n,t)+	h_5(n,t)\bigg]
\end{align*}
where $f_0, g_i$ and $h_i$ are given by 
{\allowdisplaybreaks
\begin{align*}
f_0(n,t)&=\text{sech}^9\left(\frac{\pi  t}{\beta }\right) \tanh ^{n-2}\left(\frac{\pi  t}{\beta }\right)\\
h_1(n,t)&=\frac{1}{768} (n+1) (n+2) (n+3) (n+4) \cosh \left(\frac{2 \pi  t}{\beta }\right)\\
	h_2(n,t)&=\frac{1}{960} (n+1) (n+2) (n+3) (n+4) (2 n+5) \cosh \left(\frac{4 \pi  t}{\beta }\right)\\
	h_3(n,t)&=-\frac{1}{512} (n+1) (n+2) (n+3) (n+4) \cosh \left(\frac{6 \pi  t}{\beta }\right)\\
	h_4(n,t)&=-\frac{1}{3840}(n+1) (n+2) (n+3) (n+4) (2 n+5) \cosh \left(\frac{8 \pi  t}{\beta }\right)\\
	h_5(n,t)&=\frac{1}{1536}(n+1) (n+2) (n+3) (n+4) \cosh \left(\frac{10 \pi  t}{\beta }\right)\\
	g_1(n,t)&=-\frac{(n+1) (n+2) (n+3) (2 (n-1) n-5)}{1920} \cosh \left(\frac{2 \pi t}{\beta }\right) \log \left[\cosh ^2\left(\frac{\pi t}{\beta } \right)\right]\\
	g_2(n,t)&=-\frac{(n+1) (n+2) (n+3) (n+4) \left(12 H_{n+4}-25\right) }{9216}\cosh \left(\frac{2 \pi  t}{\beta }\right)\\
	g_3(n,t)&=\frac{(n+1) (n+2) (n+3) (n+4) \left(-60 (2 n+5) H_{n+4}+274 n+625\right)}{57600} \cosh \left(\frac{4 \pi  t}{\beta }\right)\\
	g_4(n,t)&=\frac{1}{480} (n+1) (n+2) (n+3) (n (n+4)+5) \cosh \left(\frac{4 \pi  t}{\beta }\right)  \log \left[\cosh ^2\left(\frac{\pi t}{\beta } \right)\right]\\
	g_5(n,t)&=\frac{\left((n+1) (n+2) (n+3) (n+4) \left(12 H_{n+4}-25\right)\right) }{6144}\cosh \left(\frac{6 \pi  t}{\beta }\right)\\
	g_6(n,t)&=\frac{(n+1) (n+2) (n+3) (2 n-5) (2 n+3)}{3840} \cosh \left(\frac{6 \pi  t}{\beta }\right)  \log \left[\cosh ^2\left(\frac{\pi t}{\beta } \right)\right]\\
	g_7(n,t)&=-\frac{(n+1) (n+2) (n+3) (n (4 n+11)+15)}{1920}  \log \left[\cosh ^2\left(\frac{\pi t}{\beta } \right)\right]\\
	g_8(n,t)&=\frac{(n+1) (n+2) (n+3) (n+4) \left(60 (2 n+5) H_{n+4}-274 n-625\right) }{230400}\cosh \left(\frac{8 \pi  t}{\beta }\right)\\
	g_9(n,t)&=-\frac{1}{384} (n+1)^2 (n+2) (n+3) \cosh \left(\frac{8 \pi  t}{\beta }\right)  \log \left[\cosh ^2\left(\frac{\pi t}{\beta } \right)\right]\\
	g_{10}(n,t)&=\frac{(n+1) (n+2) (n+3) (n+4) \left(25-12 H_{n+4}\right) }{18432}\cosh \left(\frac{10 \pi  t}{\beta }\right)\\
	g_{11}(n,t)&=\frac{1}{768} (n+1) (n+2) (n+3) \cosh \left(\frac{10 \pi  t}{\beta }\right)  \log \left[\cosh ^2\left(\frac{\pi t}{\beta } \right)\right]\\
	g_{12}(n,t)&=\frac{(n+1) (n+2) (n+3) (n+4) \left(60 (2 n+5) H_{n+4}-274 n-625\right)}{76800}
\end{align*}}

\bibliographystyle{jhep}
\bibliography{Krylov}{}

\end{document}